\documentclass[journal]{IEEEtran}
\usepackage{amsmath}
\usepackage{amsfonts}
\usepackage{epsfig}
\usepackage{amssymb}
\usepackage{graphics}
\usepackage{}
\usepackage{float}
\usepackage{amsmath}
\usepackage{slashbox}
\usepackage[T1]{fontenc}
\usepackage{cite}
\usepackage{cleveref}
\usepackage{array}
\usepackage{caption}
\usepackage{subcaption}
\usepackage[usenames, dvipsnames]{color}

\makeatletter
\newcommand{\thickhline}{%
    \noalign {\ifnum 0=`}\fi \hrule height 1.2pt
    \futurelet \reserved@a \@xhline
}
\newcolumntype{"}{@{\hskip\tabcolsep\vrule width 1pt\hskip\tabcolsep}}
\makeatother

\begin{document}

\title{Optimization of the Belief-Propagation Algorithm for Distributed Detection by Linear Data-Fusion Techniques}

\author{\authorblockN{Younes~Abdi,~\IEEEmembership{Member,~IEEE,} and Tapani~Ristaniemi,~\IEEEmembership{Senior Member,~IEEE}}
\thanks{Y.~Abdi and T.~Ristaniemi are with the Faculty of Information Technology, University of Jyv\"askyl\"a, P.~O.~Box 35, FIN-40014, Jyv\"askyl\"a, Finland, Tel. +358 40 7214 218 \mbox{(e-mail:younes.abdi@jyu.fi, tapani.ristaniemi@jyu.fi)}. 
}\vspace{-0.25 in}
}

\maketitle

\begin{abstract}
In this paper, we investigate distributed inference schemes, over binary-valued Markov random fields, which are realized by the belief propagation (BP) algorithm. We first show that a decision variable obtained by the BP algorithm in a network of distributed agents can be approximated by a linear fusion of all the local log-likelihood ratios. The proposed approach clarifies how the BP algorithm works,  simplifies the statistical analysis of its behavior, and enables us to develop a performance optimization framework for the BP-based distributed inference systems. Next, we propose a blind learning-adaptation scheme to optimize the system performance when there is no information available a priori describing the statistical behavior of the wireless environment concerned. In addition, we propose a blind threshold adaptation method to guarantee a certain performance level in a BP-based distributed detection system. To clarify the points discussed, we design a novel linear-BP-based distributed spectrum sensing scheme for cognitive radio networks and illustrate the performance improvement obtained, over an existing BP-based detection method, via computer simulations. 
\end{abstract}


\section{Introduction}\label{sec:Intro}
\PARstart{M}{any} statistical inference problems deal with a large collection of random variables and their statistical interactions. Graphical models, a.k.a., factor graphs, which are commonly used to capture the interdependencies between correlated random variables, are known to provide a powerful framework for developing effective low-complexity inference algorithms in various applications found in diverse areas such as wireless communications, image processing, combinatorial optimization, and machine learning, see e.g.,  \cite{Kschischang01, Noorshams13, Loeliger04} and the references therein.

In many cases of practical importance, a statistical inference problem is solved by calculating marginal distribution functions which characterize the behavior of the random variables or a subset of the random variables represented within a factor graph. These marginals can be well approximated with low computational complexity by using the so-called belief propagation (BP) algorithm, a.k.a., the sum-product algorithm \cite{Mooij07, Wymeersch12}. This algorithm works as a message-passing mechanism within a network of interconnected nodes structured according to the graphical model which encodes Bayesian or Markovian interdependencies between the random variables concerned. Specifically, the messages are updated and passed across the network to enable the nodes to exchange their local inference outcomes in order to obtain a more informative global inference result taking into account the statistical interdependencies between the observations made throughout the entire network.  

In practice, there are certain obstacles in the application of the BP algorithm to concrete problems \cite{Mooij07}. We know that if the BP converges, it is not clear in general whether the results are a good approximation of the exact marginals. Even sometimes the BP does not converge and in these cases gives no approximations at all.  Due to the nonlinearity of the message-update rule in the BP algorithm, analyzing the statistical behavior of the messages (and beliefs) is certainly a challenge. The lack of understanding of the behavior of the BP algorithm makes the optimization of a BP-based inference system difficult. As explained in detail in Section \ref{sec:BP}, this message-passing algorithm works based on factorizing the joint probability distribution of the random variables, observed in the network, whose statistical behavior is described by a factor graph. However, fitting a proper factor graph into a network behavior is not an easy task in general. In \cite{Lokhov18} the major challenges in learning a graphical model from a number of statistically independent data samples are explained. 

Even though several works in distributed inference assume a factor graph to model the network statistics, they often use simple heuristic approaches to learning the parameters of such graphs in the network without directly taking into account the impact of those parameters when optimizing the system performance, see e.g., \cite{Zhang11, Li10, Penna12, Wymeersch12}.  
 In this paper, we tackle this issue by using linear data-fusion techniques \cite{Quan08, Quan09, Quan10, Taricco11} to develop a systematic framework for optimizing the performance of a BP-based distributed inference system. Specifically, through linear approximations, we simplify the message-update rule in the BP algorithm and reveal some important aspects of its behavior and demonstrate how its outcome is affected by the network configuration. We then argue that fitting a proper factor graph to the network and running a BP algorithm based on that graph is (approximately) equivalent to designing a linear data-fusion scheme. The link between the linear fusion and the BP algorithm enables us to optimize the performance of the BP-based statistical inference systems in different design scenarios. 
 
 To clarify the proposed approach, we consider the problem of distributed spectrum sensing in cognitive radio (CR) networks. In these networks, the wireless nodes perform spectrum sensing, in bands allocated to the so-called primary users (PU), in order to discover vacant parts of the radio spectrum and establish communication on those temporarily- or spatially-available spectral opportunities \cite{Akyildiz11}. In this context, CRs are considered secondary users (SU) in the sense that they have to vacate the spectrum, to avoid making any harmful interference, once the PUs are active. Hence, the spectrum sensing capability plays a crucial role in CR networks. The use of BP for distributed spectrum sensing is discussed in  \cite{Penna12, Wymeersch12} where the dependency among the heterogeneous spectral occupancy states observed by the network nodes is modeled by a binary-valued pairwise Markov random field (MRF) \cite{Wainwright08}. We use a similar graphical model and design a distributed spectrum sensing scheme based on a linear BP algorithm optimized in different design scenarios. We show that a linear message-passing algorithm obtained through linearizing the nonlinear transfer functions in a BP algorithm provides an upper bound on the performance level of the BP-based distributed detection. Consequently, by optimizing the linear version of the BP we can achieve an optimal BP algorithm for distributed detection. In particular, we offer the following contributions:  
\begin{itemize}
  \item We show that a distributed inference based on the BP algorithm can be approximated by a distributed linear data fusion scheme.  
  \item We propose a linear BP algorithm, derive its convergence conditions, and provide an effective framework for distributed optimization of its performance. 
  \item  We design a novel cooperative spectrum sensing scheme in CR networks based on the proposed framework and show that it closely obtains the optimal detection performance.
  \item We develop a blind learning-optimization scheme for performance optimization of the BP-based distributed spectrum sensing when no information is available a priori regarding the wireless environment. 
  \item We build a blind threshold adaptation mechanism, independent of the proposed linear BP, by which a certain detection performance level is guaranteed when no information is available a priori regarding the wireless environment. 
\end{itemize}
 
The rest of the paper is organized as follows. In Section \ref{sec:relatedWork}, we review  the exiting works to clarify our position with respect to the state of the art. In Section \ref{sec:BP}, we give a brief overview of the BP algorithm and demonstrate the link between the BP and distributed linear data-fusion. In Section \ref{sec:LinFus}, we discuss how to design a BP-based linear fusion scheme in both centralized and decentralized network configurations. In Section \ref{subsec:BlindDetect}, we show how to implement the proposed optimization framework when no statistical information is available a priori. In Section \ref{sec:Convergence}, we derive the convergence conditions for the proposed linear BP. We illustrate the performance of the proposed method via computer simulations in Section \ref{sec:Simulations} and provide the concluding remarks in Section \ref{sec:conclusions}.

 \section{Related Work}\label{sec:relatedWork}
 Message-passing algorithms offer effective low-complexity alternatives to standard optimization methods and have long been used in numerous signal-processing applications, such as channel decoding, image processing, CDMA systems, distributed detection, etc. These algorithms are typically used when dealing with a group of correlated random variables whose collective impact on the system performance is either difficult to analyze or is computationally expensive to be characterized numerically. A better understating of the statistical behavior of those algorithms normally leads to inference systems with better performance. To this end, several inference techniques have been proposed in the literature, based on analyzing the behavior of the messages, which use some sort of approximation. 
 
In \cite{Wiberg96, Chung01, Richardson01} Gaussian approximations for message probability density functions (pdf) are used to simplify the analysis of the BP-based decoding algorithm for low-density parity-check (LDPC) codes. The Gaussian approximations used in \cite{Wiberg96} are mainly based on empirical observations. In \cite{Chung01, Richardson01}, the code block length is assumed long enough in order to approximate the structure of the decoding subgraphs with trees. Consequently, the messages are built as sums of many independent random variables and are approximated as Gaussian random variables. For turbo codes, a one-dimensional Gaussian approximation based on extrinsic information transfer chart (EXIT chart) is used in \cite{Brink2000, Brink2000rate, El2000, El2001}. Since there is no simple formula for updating message pdf's for turbo decoding, Monte Carlo simulations are used in those works to analyze approximate trajectories for Gaussian messages under turbo decoding.
 
 Although the works in channel coding make us more motivated in using an approximate model for the BP algorithm, their results cannot be used in the distributed detection scenario considered in this paper for two strong reasons: first, we cannot rely on empirical results which describe the behavior of the BP algorithm in channel decoding schemes. We are dealing with a different application here. And, second, we do not assume that the factor graph is cycle-free.
  
In \cite{Welling04} small perturbations are inserted into a BP algorithm whose messages are formulated by first- and second-order linear approximations. Through analyzing the behavior of the perturbations and their relations to the underlying graph structure, the joint probability densities of the nodes located far away from each other are estimated. In our detection scenario, the main goal is to properly utilize the possible correlations observed between the sensing outcomes made at nodes which are possibly located close to each other. 
 
In \cite{Bayati11} the use of a linear approximate message-passing (AMP) algorithm is discussed assuming that a linear mixing problem is to be solved. Specifically, the aim in that problem is to reconstruct a sparse vector $\boldsymbol{x}$ from a small vector of linear observations $\boldsymbol{y}$ which are disturbed by linear additive noise components $\boldsymbol{w}$, i.e., $\boldsymbol{y} = \boldsymbol{A}\boldsymbol{x}+\boldsymbol{w}$. It is shown in \cite{Bayati11}, as an extension of the discussions made in \cite{Donoho09, Donoho10}, that when the mixing structure $\boldsymbol{A}$ is known, the behavior of the AMP algorithm is accurately described by a simple recursive formula referred to as \emph{state evolution}, provided that the dimensions of the problem are large. The state evolution offers a convenient way to formulate  AMP performance. \cite{Rangan2010, Rangan2011} offer a generalization of the AMP, termed GAMP, which can be used when the output channels involve arbitrary, possibly nonlinear, known operations. The BP messages are approximated as i.i.d Gaussian random variables in \cite{Rangan2010, Rangan2011}. These approximations are not justified analytically. 

It is worth noting that, the AMP and GAMP algorithms are built on a graph structure determined by $\boldsymbol{A}$. That is, in those works the main intention of the message-passing algorithm is to impose the impact of the mixing process on the collective behavior of the estimates rather than on exploiting the correlations between the elements of $\boldsymbol{x}$ (or $\boldsymbol{y}$). This can be further clarified by noting that all random variables concerned are assumed mutually independent in \cite{Rangan2010} when formulating the BP algorithm. This assumption, which cannot be justified, serves as the core of the GAMP technique by reducing the vector MAP estimation problem to a sequence of scalar MAP estimations at the inputs and outputs. In our case, such an approach in modeling the behavior of an ad-hoc sensor network would mean ignoring a major part of the correlations between the observations made by different nodes in the network.
 
 In our distributed detection scenario, if we model the state of the PU transmitters by $\boldsymbol{x}$ and the sensing outcomes by $\boldsymbol{y}$  then the graph structure depends on the network topology which is not known. In other words, $\boldsymbol{A}$ is not known in our problem since specifying its values requires the knowledge of the network topology in detail. Specifically, $\boldsymbol{A}$ depends on the channel coefficients experienced at different locations throughout the entire network. The average channel power gains, which specify the average received SNR values, can be obtained if either the PU transmit power levels are known or the sensor locations with respect to the PU transmitters are available. We do not know the sensor locations nor do we know the PU transmit power levels. Hence, in a practically-appealing detection scenario, we cannot make such assumptions about the network topology or about the behavior of the sources of the signals which are to be detected. Moreover, the AMP algorithm in \cite{Bayati11} assumes that the probability distribution of $\boldsymbol{x}$ is known. In contrast, we do not assume any prior knowledge about the behavior of the PU transmitters. 
 
We overcome these challenges by assuming Markovian dependencies between the random variables as in  \cite{Penna12, Wymeersch12}. Specifically, we use an MRF to model the statistical behavior of the network and design a BP algorithm with the aim of fitting a pairwise MRF to the network behavior. A pairwise MRF models the joint probability distribution of the random variables of interest in terms of the pairwise correlations observed in the network. These correlations are easy to obtain in practice. In addition, pairwise MRFs fit well into the commonly-used ad-hoc network configurations in which major network functionalities are conducted through pairwise i.e., one-hop, links between the nodes located close to each other. This design method is based on the common assumption that nodes located close enough to each other for one-hop communication, experience some levels of correlation between their sensor outcomes.
  
In Section \ref{subsec:BlindDetect}, we show how to blindly estimate the required statistics to characterize the radio environment according to the MRF structure discussed. One might argue that the blind estimation proposed in this paper provides the (unknown) mixing structure that we need in the AMP algorithm. This is not the case since the analysis in \cite{Bayati11} is proposed for systems with $\boldsymbol{A}$ composed of i.i.d entries. Due to the inherent spatiotemporal diversity in a network of distributed agents, we do not view the mixing coefficients as i.i.d random variables. Here we are dealing with correlations observed both in time and in space. 

 Finally, note that the work in this paper has some similarities, in its overall structure to the works in channel coding which jointly realize codeword estimation with channel parameter estimation, see, e.g., \cite{Toto10}. This joint estimation approach is used in BP-based spectrum sensing systems as well. That is, the model describing the radio environment is updated based on the results of the distributed detection. The use of BP in such systems is mainly intended to reduce the computational complexity of the joint estimation process. We do not intend to alter the overall structure of this joint estimation. What we propose here is to better implement it by better exploitation of the available resources.

\section{Belief Propagation on a Factor Graph}\label{sec:BP}
We consider a pairwise MRF defined on a graph $G = (\mathcal{V,E})$ composed of a set of vertices or nodes $\mathcal{V}$ and a set of edges $\mathcal{E} \subset \mathcal{V} \times \mathcal{V}$. Node $i \in \mathcal{V}$ corresponds to a random variable $x_i$ and edge $(i,j) \in \mathcal{E}$ connects nodes $i$ and $j$ and represents a possible correlation between random variables $x_i$ and $x_j$. This graph can be used to model the interdependency of local inference results in a network of distributed agents such as a wireless sensor network. In this network, spatially-distributed nodes exchange information with each other in order to solve a statistical inference problem. 

Let $\boldsymbol{x} = [x_1, ..., x_N]^T$ denote the vector of $N$ random variables and $ \boldsymbol{Y} = [\boldsymbol{y}_1, ..., \boldsymbol{y}_N]$ denote the vector of observations in the network where  $\boldsymbol{y}_i = [y_i(1), ..., y_i(K)]^T$ contains $K$ observation samples at node $i$. An MRF defined on factor graph $G$ is used to factorize the a posteriori distribution function $p(\boldsymbol{x}\vert \boldsymbol{Y}) $ into single-variable and pairwise terms, i.e.,    
\begin{equation}\label{p(x|y)} 
p(\boldsymbol{x}\vert \boldsymbol{Y}) \propto  \prod_{n=1}^{N} \phi_{n}(x_{n}) \prod_{(i,j) \in {\cal E}} \psi_{ij}(x_i, x_j) 
\end{equation}
where $\propto$ denotes proportionality up to a multiplicative constant. The product over couples $(i, j)$ is only over $i < j$, for which $i \in \mathcal{N}_j$ where $\mathcal{N}_j$ refers to the set of neighbors of node $j$ in the graph, i.e., $\mathcal{N}_j \triangleq \{k : (k,j) \in \mathcal{E}\}$. Each single-variable term $\phi_n(x_n)$ captures the impact of $x_n$ in the joint distribution whereas each pairwise term $\psi_{ij}(x_i, x_j) $ represents the interdependency of $x_i$ and $x_j$ connected by an edge in the graph. Please note that, as we will see in the following, $\boldsymbol{Y}$ is implicitly included in the right side of \eqref{p(x|y)} within the structures of $\phi_{n}$'s and $\psi_{ij}$'s and also in the multiplicative constant. 

The main goal of each node, say node $i$, is to find its marginal a posteriori distribution $p(x_i \vert \boldsymbol{Y})$. Since calculating the marginal  distributions from the joint distribution function requires multiple integrations, its computational complexity grows prohibitively with the number of variables $N$. However, this goal can be achieved in a distributed low-complexity fashion by using the BP algorithm based on factorizing the joint distribution function as in \eqref{p(x|y)}. In particular, BP is a parallel message-passing algorithm where the messages sent from node $k$ to node $j$ in the network are built based on multiplying three factors together: first, the local inference result at node $k$, which is $\phi_{k}(x_{k})$, second, the correlation between $x_k$ and $x_j$ which is modeled by $\psi_{kj}(x_k, x_j) $, and third, the product of all messages received from the neighbors of node $k$ except for node $j$. This term is then summed over all values of $x_k$ to create the message. More specifically, the BP message from node $k$ to node $j$, at the $l$th iteration, is formulated as  
\begin{equation}\label{mu_kj}
\mu_{k \rightarrow j}^{(l)}(x_j) \propto \sum_{x_k} \phi_k(x_k)\psi_{kj}(x_k,x_j) \prod_{n \in \mathcal{N}_k^j}\mu_{n \rightarrow k}^{(l-1)}(x_k)
\end{equation}
 where $\mathcal{N}_k^j \triangleq \mathcal{N}_k \backslash \{j\}$ denotes all neighbors of node $k$ except for node $j$. 
 
 
 The so-called \emph{belief} of node $j$ regarding $x_j$, denoted $b_j^{(l)}(x_j)$, is formed by multiplying the local inference result at node $j$ by all the messages received from its neighbors, i.e., 
\begin{equation}\label{b_j} 
b_j^{(l)}(x_j) \propto \phi_{j}(x_j) \prod_{k \in \mathcal{N}_j}\mu_{k \rightarrow j}^{(l)} (x_j)
\end{equation}
which is used to estimate (i.e., to approximate) the desired marginal distribution, i.e., $ b_j^{(l)}(x_j) \approx p(x_j \vert \boldsymbol{Y})$. The proportionality signs in \eqref{mu_kj} and \eqref{b_j} indicate that the beliefs and messages are expressed up to a constant, which can be found by normalizing $b(\cdot)$ and $\mu(\cdot)$ so as to sum to 1.

 To represent the probability measure defined on $\boldsymbol{x}$, we adopt the commonly-used exponential model \cite{Wainwright08}, i.e., 
 \begin{equation}\label{expModel} 
p(\boldsymbol{x}) \propto  \textup{exp}\left ( \sum_{n=1}^{N} \theta_{n}x_{n} + \sum_{(i,j) \in {\cal E}} J_{ij}x_i x_j \right )
\end{equation}
Note that, when $\theta_n = 0$ and $J_{ij}$ is a constant for all $n,i,j$, this model turns into the well-known Ising model \cite{Wainwright08, Lokhov18}. 

An inference scenario of high practical importance is a binary hypothesis test in which all random variables are binary-valued, i.e., $x_i \in \{0, 1\}$ for all $i \in \mathcal{V}$. The test can be formulated as $\hat{\boldsymbol{x}} = \operatorname*{max}_{\boldsymbol{x}} p(\boldsymbol{x}\vert \boldsymbol{Y}) = \operatorname*{max}_{\boldsymbol{x}} p(\boldsymbol{Y}\vert \boldsymbol{x})p(\boldsymbol{x})$. Such an optimization, even if all the joint distributions are available, incurs prohibitive computational complexity. In practice, the test is conducted by estimating the marginal distributions which are then used in a likelihood-ratio test (LRT). That is, $\hat{x}_j = \operatorname*{max}_{x_j \in \{0,1\}} p(x_j \vert \boldsymbol{Y})$ is solved, which can be realized by evaluating the so-called likelihood ratio of $x_j$, i.e., $\hat{x}_j = 1$ if $\ln [p(x_j = 1 \vert \boldsymbol{Y})/p(x_j = 0 \vert \boldsymbol{Y})] > 0$ and $\hat{x}_j = 0$ otherwise.

   The BP algorithm is used in such an inference scenario as described in the following. Given $\boldsymbol{x}$, the local observations are commonly assumed independent from each other. Consequently, we can factorize their conditional probability distribution as $p(\boldsymbol{Y}\vert \boldsymbol{x}) = \prod_{n=1}^{N} p(\boldsymbol{y}_n \vert x_n) $. This model is used in \cite{Wymeersch12,Penna12,Penna2010likelihood,Penna2010} based on the fact that $\boldsymbol{y}_n$ is evaluated at node $n$, when calculating the log-likelihood ratio (LLR), solely based on the status of $x_n$. Consequently, in this detection structure $x_n$  works as a sufficient statistic for $\boldsymbol{y}_n$, i.e., $p(\boldsymbol{y}_n \vert \boldsymbol{x}) = p(\boldsymbol{y}_n \vert x_n)$. Since the a posteriori probability distribution of $\boldsymbol{x}$ can be stated as $p(\boldsymbol{x}\vert \boldsymbol{Y}) = p(\boldsymbol{Y}\vert \boldsymbol{x})p(\boldsymbol{x})/p(\boldsymbol{Y})$, from \eqref{expModel} we have
    
\begin{equation}\label{p(x|y)exp} 
p(\boldsymbol{x}\vert \boldsymbol{Y}) \propto \prod_{n=1}^{N} p(\boldsymbol{y}_n \vert x_n) e^{\theta_n x_n} \prod_{(i,j) \in \mathcal{E} } e^{J_{ij}x_ix_j}
\end{equation}
which leads to
\begin{equation}\label{mu_kjExp} 
\mu_{k \rightarrow j}^{(l)} (x_j) \propto \sum_{x_k} p(\boldsymbol{y}_k \vert x_k) e^{\theta_kx_k} e^{J_{kj}x_kx_j} \prod_{n \in \mathcal{N}_k^j}\mu_{n \rightarrow k}^{(l-1)} (x_k)
\end{equation}
and 
\begin{equation}\label{b_jExp} 
b_j^{(l)}(x_j) \propto p(\boldsymbol{y}_j \vert x_j) e^{\theta_j x_j} \prod_{k \in \mathcal{N}_j}\mu_{k \rightarrow j}^{(l)} (x_j)
\end{equation}
The proportionality sign in \eqref{p(x|y)exp} includes but is not limited to $1/p(\boldsymbol{Y})$. 
The message update rule can be expressed in terms of the belief and message log-likelihood ratios (LLRs) defined, respectively, as 
\begin{align}
\lambda_j^{(l)} &\triangleq \ln \frac{b_j^{(l)}(x_j = 1)}{b_j^{(l)}(x_j = 0)} \label{lambda_j} \\
m_{k \rightarrow j}^{(l)} &\triangleq \ln \frac{\mu_{k \rightarrow j}^{(l)} (x_j = 1)}{\mu_{k \rightarrow j}^{(l)} (x_j = 0)} \label{m_kj} 
\end{align}
where $\lambda_j^{(l)}$ can be used as an estimation of the likelihood ratio of $x_j$ and $m_{k \rightarrow j}^{(l)}$ can be used as the message sent from node $k$ to node $j$. 
By this change of variables, we have\footnote{In \eqref{m_kj_2}, we have replaced $2\theta_k + \gamma_k$ by $\gamma_k$ and $2J_{kj}$ by $J_{kj}$ since $2\theta_k$ can be merged into $\gamma_k$ and $2$  can be merged into $J_{ij}$ with no impact on the rest of the analysis.} 
\begin{align}
m_{k\rightarrow j}^{(l)} &= S\left(J_{kj},~ \gamma_k +\sum_{n\in {\cal N}_{k}^j}m_{n\rightarrow k}^{(l-1)}\right) \label{m_kj_2} \\
\lambda_j^{(l)} &= \gamma_j +\sum_{k\in {\cal N}_{j}}m_{k\rightarrow j}^{(l)} \label{lambda_j_2} 
\end{align}
where $S(a, b)\triangleq \ln{1+e^{a+b} \over e^{a}+e^{b}}$ while $\gamma_j \triangleq \ln \frac{p(\boldsymbol{y}_j  \vert x_j= 1 )}{p(\boldsymbol{y}_j \vert x_j= 0 )}$
denotes the LLR of the local observations at node $j$. We refer to $\gamma_j$ as the local sensing outcome at node $j$. After $l^*$ iterations, $\lambda_j^{(l^*)}$ is used at node $j$ as a decision variable to perform an LRT, i.e., $\hat{x}_j = 1$ if $\lambda_j^{(l^*)} \ge \tau_j$ and $\hat{x}_j = 0$ otherwise. 


In order to properly determine the threshold  and to formulate the system performance in terms of the commonly used metrics, the probability distribution of $\lambda_j$ is needed. This is a major challenge since, first, the probability distribution of $\gamma_i$'s are not known a priori. And second, even if they are known somehow, the nonlinearity of the message-update rule in  \eqref{m_kj_2} does not allow us to find the distribution of $\lambda_j$ and the resulting performance in closed form. In \cite{Penna12}, a BP-based distributed detection method is introduced. However, due to these challenges, the detection threshold is determined through a heuristic approach and the system performance optimization is not discussed. In this paper we build an analytical framework to address these issues. 

\begin{figure}[]
	\centering 
  \includegraphics[scale=0.31]{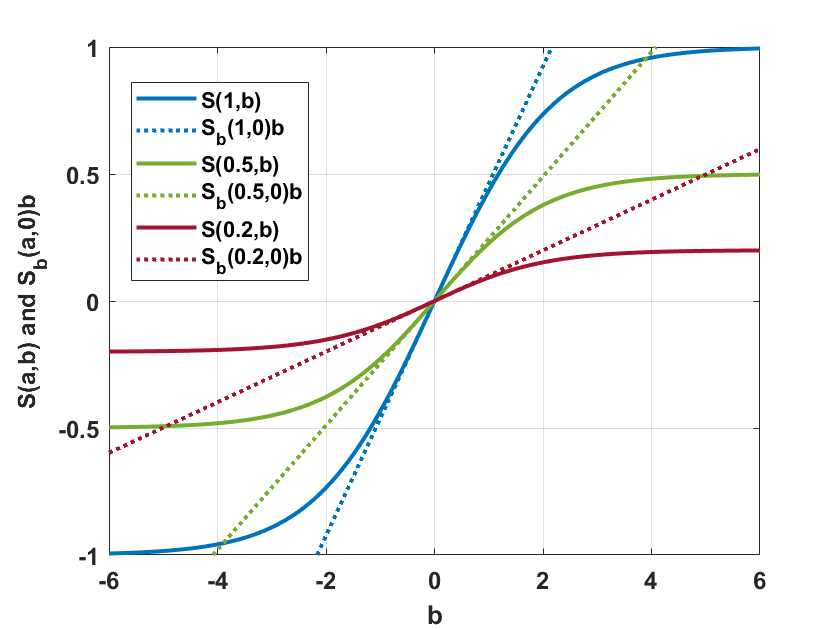} 
  \caption{Nonlinear transformation $S(a,b)$ along with its linear approximation $S_b(a,0)b$. We replace $S$, which  acts as a sigmoid function, with a linear transfer function. }
  \label{fig1}
\end{figure}

  Eq. \eqref{lambda_j_2} shows that the decision variable at node $j$, i.e., $\lambda_j$, is built by the local LLR obtained at node $j$ plus a linear combination of the messages received from the neighboring nodes. The messages, however, pass through a nonlinear transformation $S$, see Fig. \ref{fig1}. In \eqref{m_kj_2} and \eqref{lambda_j_2} we clearly see how the nonlinearity of the BP algorithm transforms the likelihood values in the factor graph.  

 A closer look at \eqref{m_kj_2} reveals that $S(a,b)$ plays a role similar to the role of the sigmoid activation functions typically used in artificial neural networks \cite{Gurney97}. In certain neural-network structures, each neuron generates an outcome by applying a bounded nonlinear transformation on the sum of its inputs received from other neurons in the network. This transfer function keeps the network stable by smoothening the neuron outcomes as well as by restricting them to reside between an upper and a lower bound. Consequently, the signals passed from one neuron to another become less sensitive to sharp variations in the inputs and the mapping learned by the network does not result in unbounded values for the variables of interest.

We propose to replace the nonlinear functions in the BP algorithm by linear ones. However, we design the proposed linear transformations in the network while taking into account the convergence condition of the distributed message-passing algorithm obtained. Having linear message-update rules in the BP algorithm enables us to analyze the network dynamics and develop effective optimization techniques to enhance system performance. 

We realize the proposed linearization by approximating the message transfer function $S$. Specifically, we use the first-order Taylor series expansion of $S(a,b)$ with respect to its second argument, i.e., $S(a,b) \approx S_b(a,0)b$, where $S_b(a,b) \triangleq \frac{\partial {S(a,b)}}{\partial b}$. This leads to a linear message-update rule as 
\begin{equation}\label{m_kj_linear} 
m_{k \rightarrow j}^{(l)} \approx c_{jk} \left ( \gamma_k + \sum_{n \in \mathcal{N}_k^j}m_{n \rightarrow k}^{(l-1)}\right ) 
\end{equation}
where $c_{jk} = \frac{(e^{2J_{kj}}-1)}{(1+e^{J_{kj}})^2}$.

 By applying this approximation on \eqref{m_kj_2} and \eqref{lambda_j_2}, while iterating the resulting linear message-update rule, we obtain  
\begin{align}\label{lambda_j_linear} 
\lambda_j  \approx \gamma_j &+ \sum_{k \in \mathcal{N}_j}c_{jk}\gamma_k + \sum_{k \in \mathcal{N}_j}\sum_{n \in \mathcal{N}_k^j}c_{jk}c_{kn} \gamma_n  \nonumber\\
  &+ \sum_{k \in \mathcal{N}_j}\sum_{n \in \mathcal{N}_k^j}\sum_{m \in \mathcal{N}_n^k}c_{jk}c_{kn}c_{nm} \gamma_m + ... 
\end{align}
which shows that, given enough time, the overall result of the BP algorithm at a node is  obtained, approximately, by a linear combination of all the local LLRs in the network. Note that we do not need the iteration index $(l)$ in this expression since \eqref{lambda_j_linear} denotes the final result of the iterations. 

Let $(V, \| \cdot \|)$ denote a normed space where $f : V \rightarrow V$ is defined and let $\boldsymbol{m}^{(l)}$ denote a vector containing all the messages at iteration $l$. A general message-passing algorithm is defined by $\boldsymbol{m}^{(l)} = f^l(\boldsymbol{m}^{(0)})$ where $f$ indicates the message-update rule. Convergence of such an iteration to a so-called \emph{fixed point} means that the sequence $\boldsymbol{m}, f(\boldsymbol{m}), f^2(\boldsymbol{m}), ...$ obtained by iterating $f$ converges to a unique fixed point for any $\boldsymbol{m} \in V$. Consequently, $\lambda_j$ in \eqref{lambda_j_linear} denotes (the decision variable made by) the fixed point of the linear BP algorithm in \eqref{m_kj_linear} and an approximation of (the decision variable made by) the fixed point attained via the BP in \eqref{m_kj_2}.

The linearization in \eqref{m_kj_linear} can be interpreted in two ways. One can see it as a linear approximation to the message-update rule in \eqref{m_kj_2}. Such an interpretation is valid since when $\gamma_i$'s are Gaussian (which is a common practical case), one can normalize them without affecting the detection result. Specifically, when the decision variable follows a Gaussian distribution, for a constant false-alarm probability the probability of detection does not change by normalizing the decision variable. Consequently, by proper normalization of $\gamma_i$'s, the messages can be made small enough to experience a linear behavior in the transfer functions. 

Alternatively, one can opt not to normalize the local sensing outcomes and to consider \eqref{m_kj_linear} as a new message-update rule which is obtained directly by replacing the nonlinear transfer functions in the network by linear ones. Note that when the message values are too large to fall within the linear part of the transfer functions, the saturation behavior in $S$ acts as a thresholding process leading to a so-called hard fusion of the local LLRs. By using the linear transfer function in \eqref{m_kj_linear}, the hard fusion rule is replaced by a soft one in which the sensing outcomes are directly combined without passing through that threshold. Since a soft fusion method outperforms a hard fusion one in general, one can expect a better overall detection result obtained by this linearization. A hard fusion scheme refers to a combination of binary-valued variables obtained by comparing the local sensing outcomes against (local) decision thresholds. See \cite{Chaudhari12} for a comparison between the hard and soft decision methods.  

\emph{Remark 1}: Regardless of how \eqref{m_kj_linear} is interpreted, since the sigmoid functions are now removed, the linear transfer functions have to be designed properly to guarantee the convergence of the new message-update rule.  We will discuss the convergence conditions in Section \ref{sec:Convergence}. 

\emph{Remark 2}: We have configured the linear combination in \eqref{lambda_j_linear} such that it reveals the effect of the network topology on how the fusion coefficients are arranged. Specifically, for the one-hop neighbors of node $j$ we have one coefficient $c_{jk}$ affecting the local likelihoods received, for the two-hop neighbors we have two coefficients $c_{jk}c_{kn}$ and so on.  We will use this observation in Section \ref{subsec:LinFusDecent} to develop a decentralized distributed inference method. 

From \eqref{lambda_j_linear} it is clear that, determining $c_{jk}$'s for all $(k,j) \in \mathcal{E}$ in the network, based on a set of performance criteria, can be viewed as determining the values of $J_{kj}$'s in the MRF. In  \cite{Penna12, Wymeersch12}, the BP is designed and analyzed without directly considering the problem of fitting an MRF to the network behavior. Specifically, in those works, $J_{kj}$'s are set by an empirical estimation of the correlations between the neighboring nodes in a window of $T$ time slots, i.e., 
\begin{equation}\label{lrnFact} 
J_{kj} \triangleq \frac{\zeta}{T}\sum_{t=1}^{T} \boldsymbol{1}\{\hat{x}_j(t) = \hat{x}_k(t)\}-\boldsymbol{1}\{\hat{x}_j(t) \neq \hat{x}_k(t)\}
\end{equation}
 where $\zeta$, referred to as the learning factor in this paper, is considered as a small constant and $\boldsymbol{1}\{A\}$ equals $1$ if $A$ is true and $0$ otherwise. $T$ denotes the number of samples used in this training process. We will see in Section \ref{sec:Simulations} why $\zeta$ should be small. 
 
 Such an approach does not directly take into account the impact of the learned factor graph on the system performance. In this paper, however, we argue that $J_{kj}$'s provide extra degrees of freedom which can be used to optimize the system performance. 
Specifically, \eqref{m_kj_2} shows that $J_{kj}$'s determine the behavior of the transfer functions applied on the BP messages. Consequently, optimizing $J_{kj}$'s can be viewed as finding an optimal set of transfer functions for the BP algorithm. Since we can normalize $\gamma_i$'s without affecting the local detection performance, the proposed linear transfer functions are included in the feasibility region of this optimization. And, since a soft fusion method is better than a fusion method which clips the local LLRs, the performance of a BP iteration with the proposed linear transfer functions gives an upper bound on the detection performance of the BP algorithm defined by \eqref{m_kj_2}. Hence, by using linear transfer functions and by optimizing the resulting message-passing algorithm, we are optimizing the parameters of the BP algorithm in a distributed detection scenario.

Alternatively, note that the proposed linear BP algorithm defined by \eqref{m_kj_linear}  is the same as the BP algorithm with \eqref{m_kj_2} where $J_{kj}$'s are large enough. And, $J_{kj}$'s are the degrees of freedom in this design. Consequently, from an optimization perspective, there is no difference between the two message-passing algorithms. 

\section{Linear Fusion Based on Belief Propagation}\label{sec:LinFus}
In this section, we use the link between the BP and linear data-fusion to design distributed inference methods in both \emph{centralized} and \emph{decentralized} network configurations. In a centralized setting  \cite{Akyildiz11}, there is a special node in the network, referred to as the fusion center (FC), which collects data from the sensing nodes, coordinates the distributed inference, and builds the overall inference outcome based on the collected data. In a decentralized (i.e., ad-hoc) network, however, there is no FC and the statistical inference is realized through the so-called peer-to-peer communication between the nodes.

 \subsection{Centralized Distributed Inference} \label{subsec:LinFusCentral}
Eq. \eqref{lambda_j_linear} can compactly be stated as $\lambda_j = \sum_{i=1}^{N} a_{ji}\gamma_i$
or, in matrix form, as 
\begin{equation}\label{lambdaVect} 
\boldsymbol{\lambda} = \boldsymbol{A}^T \boldsymbol{\gamma}
\end{equation}
where $\boldsymbol{\lambda} = [\lambda_1 , ..., \lambda_N]^T$, $\boldsymbol{\gamma} = [\gamma_1 , ..., \gamma_N]^T$, and $\boldsymbol{A} = [\boldsymbol{a}_1, \boldsymbol{a}_2, ..., \boldsymbol{a}_N]$ where $\boldsymbol{a}_j = [a_{j1}, ..., a_{jN}]^T$. 
Assuming $U(\boldsymbol{\lambda})$ as the objective function corresponding to a certain application and $C_i(\boldsymbol{\lambda})$ as one of $N_C$ constraint functions indicating the requirements to be met by the system design, the optimal linear fusion in the network can be obtained by 
\begin{gather*}\label{P1}
\operatorname*{max} U(\boldsymbol{\lambda}), ~~
\begin{matrix}
\textup{s.t.}, & C_i(\boldsymbol{\lambda}) \le \alpha_i, & i = 1, ..., N_C 
\end{matrix} \tag{P1} 
\end{gather*}

This is a generic linear fusion problem whose structure depends on the application at hand. We further clarify our points by considering a distributed spectrum sensing scenario where each node, say node $j$, performs a binary hypothesis test on the decision variable $\lambda_j$ to decide the availability of the radio spectrum. 

In this scenario, $x_j$ is a binary-valued random variable representing the occupancy state of the radio spectrum where node $j$ is located. Let's say, $x_j = 1$ indicates that the spectrum band sensed by node $j$ is occupied whereas $x_j = 0$ indicates the availability of that band for the secondary use.  
The signal samples used to form the test statistics at node $j$ can be expressed as
\begin{equation}\label{y_j}
y_j(k) = \sum_{n}u_n h_{jn} s_n(k) + \nu_j(k)
\end{equation}
where $\nu_j(k) \sim \mathcal{CN}(0, \sigma_{\nu}^2)$ denotes a zero-mean complex-valued Gaussian random variable which models the additive white noise at node $j$ while $s_n(k)$, also a zero-mean complex-valued Gaussian random variable, represents the signal transmitted by the $n$th PU transmitter with $E[|s_n(k)|^2] = \sigma_{s}^2$.  The channel gain from the $n$th PU transmitter to node $j$, denoted $h_{jn}$, is assumed constant during the sensing time interval. $u_n$ denotes a binary random variable which represents the state of the $n$th PU transmitter, i.e., the $n$th PU transmitter is on when $u_n = 1$ and is off otherwise. $u_n$ is also assumed constant during the sensing time interval.  According to this model, $x_j = 0$ means that no PU signal is received at node $j$, i.e., $u_n = 0$ for all $n$ and the spectrum is available for the secondary use. 

Given the state of $x_j$, if we assume as in \cite{Penna12} that $\boldsymbol{y}_j$ is a vector of i.i.d. complex Gaussian signal samples, then  we have 
\begin{equation}\label{gamma_j}
\gamma_j = \frac{-N}{2}\ln(1+\rho_j) + \frac{\rho_j}{1+\rho_j} \frac{\left \| \boldsymbol{y_j} \right \|^2}{2\sigma_{\nu}^2}
\end{equation}
where $\rho_j \triangleq |h_j|^2\sigma_{s}^2 / \sigma_{\nu}^2$ denotes the received signal-to-noise ratio (SNR) at node $j$. Therefore, one needs to know the channel (power) gain, the transmitted signal power, and the noise level to obtain the probability distribution of $\gamma_j$. 

In order to keep the resulting false-alarm rate below a predefined constraint $\alpha$, in \cite{Penna12} $\gamma_j$ in \eqref{gamma_j} is replaced by 
\begin{equation}\label{Penna'sED}
\gamma_j \triangleq  \frac{1}{K} \| \boldsymbol{y}_j \|^2 - \tau_0
\end{equation}
where $\tau_0$ is set such that $\textup{Pr}\{\gamma_j > 0 \vert x_j = 0\} = \alpha$, which leads to
\begin{equation} \label{tau_0}
\tau_0 = \sigma_\nu^2 \left (1+\sqrt{\frac{2}{K}}Q^{-1}(\alpha) \right )
\end{equation} 

This approach is practically appealing in the sense that, $\tau_0$ can be simply calculated in terms of the noise level and without the need for the channel gain and the transmitted power level.


 The proposed structure remains intact if we use $\gamma_j \triangleq \| \boldsymbol{y}_j \|^2/K$ instead of \eqref{gamma_j}. All other variables in \eqref{gamma_j} along with $\tau_0$ in \eqref{Penna'sED} can be absorbed either into the coefficients in \eqref{lambda_j_linear} or into the detection threshold $\tau_j$. Consequently, we can justify here why the BP algorithm works well in \cite{Penna12} where energy detection is adopted as the local sensing method. 
   As we show in Section \ref{sec:Simulations}, $\tau_0$ in \eqref{Penna'sED}   works well only for small values of $\alpha$. We propose a new threshold adaptation scheme in Section \ref{subsec:Calibration}, which can always guarantee the false-alarm rate of the BP algorithm fall below a predefined upper bound without the need for any statistics other than the local noise level. 

  
The performance of the binary hypothesis test discussed is formulated by the so-called false-alarm and detection probabilities defined, respectively, at node $j$ as  $P_\textup{f}^{(j)} \triangleq \textup{Pr} \{\lambda_j \ge \tau_j \vert x_j = 0\}$ and $P_\textup{d}^{(j)} \triangleq \textup{Pr} \{\lambda_j \ge \tau_j \vert x_j = 1\}$. 
The proposed transfer functions are linear and, given the status of $x_j$'s, our $\gamma_i$'s are Gaussian random variables. Consequently, we can express the proposed system false-alarm and detection probabilities in closed form. These probabilities at node $j$ depend not only on the state of $x_j$ but also, in general, on the state of all other $x_i$'s being sensed throughout the entire network. We take these interdependencies into account by the total probability theorem. Specifically, at node $j$ and for $v = 0,1$ we have 
\begin{align} 
& g_{j}(\tau_j,v) \triangleq \textup{Pr}\{\lambda_j > \tau_j \vert x_j = v\} \nonumber \\
 &= \sum_{\boldsymbol{b} \in \{0,1\}^{N-1}}\textup{Pr}\{\lambda_j > \tau_j \vert \boldsymbol{x}_{(j)} = \boldsymbol{b}, x_j = v\}p_{\boldsymbol{x}_{(j)} |x_j}(\boldsymbol{b}|v) \nonumber\\
 &= \sum_{\boldsymbol{b} \in \{0,1\}^{N-1}} Q \left(\frac{\tau_j - \eta_{j,v}(\boldsymbol{b})}{\sigma_{j,v}(\boldsymbol{b})} \right )p_{\boldsymbol{x}_{(j)} |x_j}(\boldsymbol{b}|v) \label{g}
\end{align}
where $\boldsymbol{x}_{(j)} \triangleq [x_1,x_2, ..., x_{j-1}, x_{j+1}, x_{j+2}, ..., x_N]$, $p_{\boldsymbol{x}_{(j)}|x_j}(\boldsymbol{b}|v) \triangleq \textup{Pr}\{\boldsymbol{x}_{(j)} = \boldsymbol{b} \vert x_j = v\}$, and for $v=0,1$, $\eta_{j,v}(\boldsymbol{b}) \triangleq E[\lambda_j \vert \boldsymbol{x}_{(j)}= \boldsymbol{b}, x_j = v]$ and $\sigma_{j,v}^2(\boldsymbol{b}) \triangleq \textup{Var}[\lambda_j \vert \boldsymbol{x}_{(j)} = \boldsymbol{b}, x_j = v]$. $Q(x) \triangleq \int_{x}^{\infty} \frac{1}{\sqrt{2\pi}} e^{-z^2/2} dz$ is the so-called $Q$-function. 
Note that $\boldsymbol{x}_{(j)}$ contains all $x_i$'s except for $x_j$. It is clear that $P_\textup{f}^{(j)} = g_j(\tau_j,0)$ and $P_\textup{d}^{(j)} = g_j(\tau_j,1)$. By solving $g_j(\tau_j,0) = \alpha$ or $g_j(\tau_j,1) = \beta$ we obtain a value for $\tau_j$ which guarantees the false-alarm or detection probability at node $j$ be, respectively, equal to $\alpha$ or $\beta$. The fact that $\lambda_j$ is Gaussian comes from \eqref{lambda_j_linear}.


In a CR network, these parameters determine the secondary network throughput and the interference level caused on the PUs. Specifically, the higher the false-alarm probability the more vacant spectrum bands are mistakenly treated as occupied, leading to a loss in the network throughput. Consequently, the opportunistic spectral utilization of the $j$th band is measured by $1 - P_\textup{f}^{(j)}$. In addition, the higher the misdetection probability the higher the interference level imposed on the primary users. Hence, the interference level on the $j$th band is measured by $1 - P_\textup{d}^{(j)}$. In a rational design strategy, interference on PUs is associated with a cost.

 We collect these performance metrics regarding the entire network in vectors  $\boldsymbol{P}_\textup{f} \triangleq [P_\textup{f}^{(1)}, ..., P_\textup{f}^{(N)}]^T$ and $\boldsymbol{P}_\textup{d} \triangleq [P_\textup{d}^{(1)}, ..., P_\textup{d}^{(N)}]^T$. Consequently, we can express the network performance by the following quantities 
 \begin{eqnarray}
R(\boldsymbol{\lambda}, \boldsymbol{\tau}) \triangleq \boldsymbol{r}^T \left [ \boldsymbol{1} - \boldsymbol{P}_\textup{f}(\boldsymbol{\lambda}, \boldsymbol{\tau}) \right ]\label{EQ26} \\
I(\boldsymbol{\lambda}, \boldsymbol{\tau}) \triangleq \boldsymbol{q}^T \left [ \boldsymbol{1} - \boldsymbol{P}_\textup{d}(\boldsymbol{\lambda}, \boldsymbol{\tau})\right ] \label{EQ27} 
 \end{eqnarray}
where $\boldsymbol{1}$ denotes an all-one vector,  $R(\boldsymbol{\lambda}, \boldsymbol{\tau})$ denotes the \emph{aggregate network throughput} calculated over all spectrum bands sensed across the whole network and $I(\boldsymbol{\lambda}, \boldsymbol{\tau})$ denotes the \emph{aggregate interference} caused on the PUs operating on those bands, where  $\boldsymbol{\tau}  \triangleq [\tau_1, ..., \tau_N]^T$ and $\boldsymbol{r} \triangleq [r_1, ..., r_N]^T$ while $r_i$ denotes the throughput achieved by using the $i$th band. The cost of interference caused on  the $i$th band is denoted by $q_i$ which is included in $\boldsymbol{q} \triangleq [q_1, ..., q_N]^T$. 

Now, the optimal sensing performance is obtained by maximizing the aggregate system throughput subject to a constraint on the aggregate interference as well as the per channel false-alarm and misdetection constraints, i.e., 
\begin{gather*}\label{P2}
 \operatorname*{max}_{\boldsymbol{A},\boldsymbol{\tau} } R(\boldsymbol{\lambda}, \boldsymbol{\tau}) \tag{P2} \\ 
\begin{matrix}
\textup{s.t.} & I(\boldsymbol{\lambda}, \boldsymbol{\tau}) \le I_0, & \mathbf{P}_\textup{f}(\boldsymbol{\lambda}, \boldsymbol{\tau}) \le \boldsymbol{\alpha}, & \mathbf{P}_\textup{d}(\boldsymbol{\lambda}, \boldsymbol{\tau}) \ge \boldsymbol{\beta}
\end{matrix}
\end{gather*}
where $\boldsymbol{\alpha} = [\alpha_1, ..., \alpha_N]^T$ and $\boldsymbol{\beta} = [\beta_1, ..., \beta_N]^T$ impose per-channel constraints on the false-alarm and misdetection probabilities. The \emph{sequential optimization} method in \cite{Quan09}, which maximizes the so-called \emph{deflection coefficient} \cite{Kay93} of the detector, can be used to solve \eqref{P2} for a good suboptimal performance. 

The deflection coefficient associated with $\lambda_j$ is defined as
\begin{equation}\label{DC}
\Delta_j \triangleq \frac{E[\lambda_j \vert x_j = 1] - E[\lambda_j \vert x_j = 0]}{\sqrt{\textup{Var}[\lambda_j \vert x_j = 0]}}
\end{equation}
where, from \eqref{lambdaVect}, $E[\lambda_j \vert x_j ]= \sum_{i = 1}^{N}a_{ji} E[\gamma_i \vert x_j ]$ and $\textup{Var}[\lambda_j  \vert x_j ] = \sum_{i = 1}^{N} \sum_{k = 1}^{N}a_{ji} a_{jk} \textup{cov}(\gamma_i, \gamma_k  \vert x_j )$. Hence, to maximize the deflection coefficients in the network, we only need  $E[\gamma_i \vert x_j]$ and $\textup{cov}(\gamma_i, \gamma_k \vert x_j)$ for all $j \in \mathcal{V}$ and $i,k \in \mathcal{E}$. These statistics are clearly easier to obtain than $\eta_{j,v}(\boldsymbol{b})$, $\sigma_{j,v}^2(\boldsymbol{b})$, and $p_{\boldsymbol{x}_{(j)}|x_j}(\boldsymbol{b}|v) $  in \eqref{g}.

   Availability of the statistics discussed may require a centralized configuration in which an FC collects data from the nodes. As explained in \cite[Sec. II-C]{Paysarvi-Hoseini11}, the standard preambles or synchronization symbols, which typically reside within the PU signals, can facilitate the estimation of the desired statistics in a centralized setting. In Section \ref{subsec:LinFusDecent}, we show how to optimize the system performance in a decentralized network and when those statistics are not available a priori. 
  
\emph{Remark 3}: Due to the nonlinearity of the message update rule in \eqref{m_kj_2}, the system false-alarm and detection probabilities are not available in closed form when the BP algorithm is used and, therefore, the performance optimization in \eqref{P2} or even guaranteeing a predefined level of performance in a BP-based system analytically is a very challenging task, if possible. 

\subsection{Decentralized Distributed Inference}\label{subsec:LinFusDecent}
 Message-passing algorithms are of special interest in the decentralized network configurations where there is no FC, the communication between nodes is limited to one-hop neighbors, and, typically, there are no statistics available a priori regarding the wireless environment.  
 
In order to build a decentralized distributed inference, we make three particular observations here. Firstly, note that $|c_{jk}| < 1$ for all $(j,k)\in \mathcal{E}$. Moreover, as shown in Section \ref{sec:UTRW-BP}, we can use $\rho c_{jk}$'s as the fusion coefficients, where $0<\rho<1$, without affecting the detection probability obtained. Scaling down the coefficients is an effective method to mitigate the impact of graph cycles on the system performance.

Secondly, our MRF in \eqref{p(x|y)} captures the correlations between the random variables by a set of pairwise links. In particular, if node $n$ is connected to node $j$ through node $k$, the correlation between $x_n$ and $x_j$ is accounted for in \eqref{p(x|y)} by two factors $\psi_{k,n}(x_k,x_n)$ and $\psi_{jk}(x_j,x_k)$ multiplied together within $p(\boldsymbol{x} \vert \boldsymbol{Y})$. A similar process is observed in \eqref{lambda_j_linear} where the impact of $\gamma_n$ on $\lambda_j$, which depends on the correlation between $x_n$ and $x_j$, is determined by multiplying two factors $c_{kn}$ and $c_{jk}$ corresponding, respectively, to the link from node $n$ to node $k$ and the link from node $k$ to node $j$.

Thirdly, the fusion process in \eqref{lambda_j_linear} is inherently symmetric. That is, each node performs the same local fusion on the data received from its neighbors and passes the result on along the path towards node $j$. We use this symmetric structure, which is clarified in \eqref{eq:floatingequation}, to distribute the load of the desired optimization across all nodes in the network. Specifically, we assign each node the task of optimizing its own local fusion process. In this manner, each node has to deal with a few variables and with the statistics received from its one-hope neighbors only. The fact that $|c_{jk}| < 1$ makes this collection of localized optimizations a reliable distributed approach to the system performance optimization. The reason is that $\gamma_i$'s received from neighbors with longer distance are multiplied by higher-order terms in the data-fusion process. Consequently, the system inherently favors, at each node, the data received from the close neighbors over the data from the more-distant ones. 
\begin{figure*}[t!]
\normalsize
\begin{IEEEeqnarray}{rCl}
\lambda_j  \approx \underbrace{\gamma_j + \sum_{k \in \mathcal{N}_j}c_{jk}}_{\text{optimized by node \emph{j}}} \Bigg ( \underbrace{\gamma_k + \sum_{n \in \mathcal{N}_k^j}c_{kn}}_{\text{optimized by node \emph{k}}} \Bigg (\underbrace{\gamma_n  
  + \sum_{m \in \mathcal{N}_n^k}c_{nm}}_{\text{optimized by node \emph{n}}} \Bigg (\gamma_m + ... 
\label{eq:floatingequation}
\end{IEEEeqnarray}
\hrulefill
\vspace*{0pt}
\end{figure*}

Therefore, firstly, 
we further approximate \eqref{lambda_j_linear} by keeping the first-order terms,
i.e., 
\begin{equation}\label{lambda_j_neighbors}
\lambda_j  \approx \gamma_j + \sum_{k \in \mathcal{N}_j}c_{jk}\gamma_k 
\end{equation}

We use $\lambda_j \approx  \sum_{k \in \mathcal{M}_j}c_{jk}\gamma_k $, where $\mathcal{M}_j \triangleq \{j\} \cup \mathcal{N}_j$, when optimizing the coefficients. In the vector form we have $\lambda_j \approx \mathbf{c}_j^T \boldsymbol{\gamma}_j$ where $\mathbf{c}_j$ contains all $c_{jk}$'s such that $k \in \mathcal{M}_j$ and $\boldsymbol{\gamma}_j$ contains the local likelihood ratios at node $j$ and at all its neighbors. 

\emph{Remark 4}: The approximation in \eqref{lambda_j_neighbors} is only used to realize the optimization of $c_{jk}$'s in a decentralized configuration. The resulting coefficients are then used in a linear message-update rule as in \eqref{m_kj_linear}. Hence, the overall impact of the distributed linear BP obtained is still approximated by \eqref{lambda_j_linear}. 

Secondly, we optimize $\mathbf{c}_j$ by considering the one-hop neighbors of node $j$. In this manner, the correlation between $x_j$ and $x_k$ is captured in $c_{jk}$ while the correlation between $x_k$ and $x_n$ is accounted for by $c_{kn}$. Hence, both of the correlations concerned are  taken into account in the system design by the multiplication $c_{jk}c_{kn}$ in \eqref{lambda_j_linear} while each node sees its immediate neighbors only. This optimization can be performed conveniently in a distributed fashion since each node has to deal with only a few variables. 

Contrary to the linear fusion method discussed in Section  \ref{subsec:LinFusCentral}, here we have an iterative message-passing algorithm whose convergence needs to be guaranteed
. In Section \ref{sec:Convergence}, we show that the convergence is guaranteed when we have $|c_{j,k}| < \frac{1}{\operatorname*{max}_{n} |\mathcal{N}_n|-1}, \forall (j,k) \in \mathcal{E}$ which is a convex constraint on the coefficients. Therefore, the proposed decentralized linear fusion scheme is obtained by solving the following optimization at node $j$, $j = 1, ..., N$, 
\begin{gather*}\label{P3}
\operatorname*{max}_{\boldsymbol{c}_j, \tau_j } U_j(\lambda_j) \tag{P3} \\
 \begin{matrix}
\textup{s.t.}, & C_{ji}(\lambda_j) \le \alpha_{ji}, ~~ \textup{for}~ i = 1, ..., N_{C_j} \\
& |c_{j,k}| < \frac{1}{\operatorname*{max}_{n} |\mathcal{N}_n|-1}, \forall k \in \mathcal{N}_j
\end{matrix} 
\end{gather*}
where $U_j(\lambda_j)$, $C_{ji}(\lambda_j)$, and $N_{C_j}$ denote the application-specific local objective function, the local constraint functions, and the number of constraints at node $j$, respectively. The major advantage of this optimization is that the local objective and constraint functions are calculated in terms of the statistics of $\lambda_j$ and these statistics are now conveniently obtained due to the linearity of the proposed message update rule. 


It is worth noting that, considering the overall system performance, the use of \eqref{lambda_j_neighbors} does not mean that the impact of the higher-order terms on $\lambda_j$ is ignored. Instead, the task of optimizing those terms are assigned to the other nodes in the network. It is clear that by this approximation we achieve a suboptimal solution. However, this solution distributes the optimization load well over the entire network by exploiting the symmetry in the overall data-fusion process achieved. Moreover, the use of \eqref{lambda_j_neighbors} is supported by the observations made in \cite{Chaudhari12} which investigates the performance of different fusion methods operating on erroneous data. In particular, it is shown in \cite{Chaudhari12} that a soft fusion scheme is rather robust against the errors in the local test summaries reported by the cooperating nodes. Accordingly, we consider the impact of the truncation in \eqref{lambda_j_neighbors}  as a reporting error in the network and expect the system to tolerate it significantly. Taking advantage of this robustness in the linear fusion scheme, we break down the desired optimization into a set of small local optimizations which can be conducted collectively in a decentralized distributed setting. The numerical results in Section \ref{sec:Simulations} support the effectiveness of the proposed optimization. 

In the rest of the paper, we use $c_{jk}^{\textup{BP}}$ to denote the coefficients in the linear approximation of the BP algorithm, i.e., $c_{jk}^{\textup{BP}} \triangleq (e^{2J_{kj}}-1)/(1+e^{J_{kj}})^2$, to distinguish those coefficients from the ones obtained by the proposed optimization. In addition,  we use $\mathbf{c}_j^{\textup{BP}} $ to contain all $c_{jk}^{\textup{BP}} $'s such that $k \in \mathcal{M}_j$.

In a spectrum sensing scenario, the proposed optimization is built by the well-known Neyman-Pearson hypothesis test \cite{Kay93} at each node. Specifically, at each node we aim at maximizing the detection probability subject to an upper bound on the false-alarm probability, i.e., 
\begin{gather*}\label{P4}
 \operatorname*{max}_{\boldsymbol{c}_j, \tau_j } P_{\textup{d}}^{(j)}(\lambda_j) \tag{P4} \\ 
\begin{matrix}
\textup{s.t.,} & P_{\textup{f}}^{(j)}(\lambda_j) \le \alpha_j\\
& |c_{j,k}| < \frac{1}{\operatorname*{max}_{n} |\mathcal{N}_n|-1}, \forall k \in \mathcal{N}_j
\end{matrix}
\end{gather*}
where $P_{\textup{f}}^{(j)}$ and $P_{\textup{d}}^{(j)}$ are derived, approximately, in a decentralized fashion as described in the following. 

The conditional probabilities in \eqref{g}, which govern the statistical behavior of the test summary $\lambda_j$ at node $j$, take into account any possible correlations between $x_j$ and all other $x_i$'s. Some nodes may, in fact, be located far away from node $j$ and their observations may have, if any, a little or negligible impact on the detection result at node $j$. Hence, we simplify the calculations by limiting the summation in \eqref{g} to include only the correlations between $x_j$ and $x_i$'s observed by the one-hop neighbors of node $j$. These statistics can be obtained by each node in a network with a decentralized configuration. The proposed approximation is expressed as 
\begin{align}\label{gApprox}
g_j(\tau_j,v) \approx \sum_{\boldsymbol{b} \in \{0,1\}^{\left|\mathcal{N}_j\right|}} Q \left(\frac{\tau_j - \tilde{\eta}_{j,v}(\boldsymbol{b})}{\tilde{\sigma}_{j,v}(\boldsymbol{b})} \right )p_{\tilde{\boldsymbol{x}}_{(j)} |x_j}(\boldsymbol{b}|v)
\end{align}
where $\tilde{\boldsymbol{x}}_{(j)}$ is a vector which contains $x_i$'s with $i \in \mathcal{N}_j$ while for $v = 0,1$, we have $\tilde{\eta}_{j,v}(\boldsymbol{b}) \triangleq E[\lambda_j \vert \tilde{\boldsymbol{x}}_{(j)} = \boldsymbol{b}, x_j = v]$ and $\tilde{\sigma}_{j,v}^2(\boldsymbol{b}) \triangleq \textup{Var}[\lambda_j \vert \tilde{\boldsymbol{x}}_{(j)} = \boldsymbol{b}, x_j = v]$. Note that $\left|\mathcal{N}_j\right|$ denotes the number of one-hop neighbors of node $j$ while $\tilde{\eta}_{j,v}$ and $\tilde{\sigma}_{j,v}$ denote an estimation of the first- and second-order conditional statistics of the test summary at node $j$ given the value of $x_j$ and its immediate neighbors. And, $\lambda_j$ is approximated as in \eqref{lambda_j_neighbors}. 

Based on \eqref{lambda_j_neighbors} we have $E[\lambda_j \vert x_j] \approx \sum_{k \in \mathcal{M}_j}c_{jk}E[\gamma_k \vert x_j]$ and $\textup{Var}[\lambda_j \vert x_j] \approx$ $\sum_{n \in \mathcal{M}_j}\sum_{k \in \mathcal{M}_j}c_{jn}c_{jk} \textup{cov}(\gamma_n,\gamma_k \vert x_j)$. Hence, the use of the deflection coefficient here can simplify the design even further compared to solving \eqref{P4}. Maximizing the deflection coefficient is a  classic low-complexity approach to optimal linear fusion which is considered in e.g., \cite{Quan08, Quan09, Quan10}. 

\emph{Remark 5}: Since each node is typically connected to only a few neighbors, regardless of the scale of the network, each node has to deal only with a few variables when running the proposed optimization (and, as discussed later, when estimating the required statistics). Consequently, the scalability of the proposed detection scheme does not depend on the low-complexity optimization techniques in \cite{Quan08, Quan09, Quan10}.

\subsection{Discussion}\label{subsec:limit}
The use of the BP algorithm, linear data-fusion, and MRFs in statistical inference systems cover a wide range of applications and design scenarios in the literature. However, these methods are normally investigated separately and in different contexts. In this paper, we intend to provide a more comprehensive view and shed light on the potential links between these effective tools. Therefore, we aim at presenting our ideas in an as generic of a framework as possible to help a reader possibly interested in a different application better understand the links discussed. The use of generic expressions to denote the objective and cost functions in \eqref{P1} and \eqref{P3} serves this purpose. 

However, when interpreting the results of the proposed optimization framework, it should be noted that we focus on distributed detection in a wireless network.  Consequently, some limitations may arise from the use of certain tools in this paper such as the type of the MRF model adopted or as the false-alarm and detection probabilities used to measure the system performance. Nevertheless, in the context of distributed detection, the models we use in this paper do not restrict the practicality of the design proposed. All the assumptions regarding the MRF, local sensing method, performance metrics, optimization scenarios, availability of channel statistics, network configuration, and the possibility of one-hop communications, etc. are in line with the  design scenarios widely-used in the literature. 

In fact, the proposed distributed optimization is based on the linear fusion structure in \eqref{lambda_j_linear}  and its symmetric nature as clarified in \eqref{eq:floatingequation}. The linear fusion discussed is mainly based on \eqref{p(x|y)} and \eqref{expModel} which together build a relatively generic model for a distributed binary hypothesis test. Due to the reasons discussed in Section \ref{sec:relatedWork}, modeling the joint pdf of the variables in terms of pairwise correlations as in \eqref{p(x|y)} is a commonly-used approach in distributed detection. In addition, the use of an exponential form in \eqref{expModel} to model the behavior of correlated binary-valued random variables is a popular choice since, first, it is supported by the \emph{principle of maximum entropy} \cite{Wainwright08} and,  second, when the graphical model is built in terms of products of functions, these products turn into additive decompositions in the log domain. 

The use of the false-alarm and detection probabilities, as performance metrics, only affects the structure of the objective and constraint functions discussed. As clarified in \emph{Remark 5}, when using the proposed distributed optimization, each deals with a relatively small optimization problem and, therefore, the type of the objective and constraint functions in \eqref{P4} are not of major concern. Hence, we do not lose the generality of the proposed distributed optimization by the use of these metrics. Our main goal in such formulations is to facilitate the design of the CR network since the false-alarm and detection probabilities are directly linked to the CR network throughput and the level of interference experienced by the PUs, respectively. This link is explained in Section \ref{subsec:LinFusCentral}.

\section{Blind Distributed Detection}\label{subsec:BlindDetect}
We propose the notions of \emph{adaptive linear belief propagation} and \emph{detector calibration} here to realize a near-optimal distributed detection performance when there is no information available a priori regarding the statistical behavior of the radio environment.   

\subsection{Offline Learning and Optimization}\label{subsec:OfflineBP}
As we saw in Section \ref{sec:LinFus}, in order to optimize the fusion weights in the proposed linear message-passing algorithm, we need the first- and second-order statistics of the local sensing outcomes. Since we do not assume any prior knowledge about the channel statistics or the behavior of the PUs, we have to estimate those parameters based on the detection outcome.  This is a challenging task since the detection outcome itself depends on those estimates. Erroneous detection outcomes lead to erroneous estimates of the desired statistics which, in turn, deteriorate the detection results. We overcome this challenge by implanting an adaptation process within the linear BP algorithm, which is run offline on a set of data samples collected and processed for channel estimation and detector optimization purposes. This adaptation is realized by an offline linear BP algorithm whose outcome is a set of near-optimal coefficients used in the main linear BP algorithm which is responsible for spectrum sensing in real time. 

In a blind detection scheme the only source of information is the detector outcome. Hence, we have to use $E[\lambda_j \vert \hat{x}_j ]$ and $\textup{cov}(\gamma_i, \gamma_k  \vert \hat{x}_j )$ to estimate  $E[\gamma_i \vert x_j ]$ and $\textup{cov}(\gamma_i, \gamma_k  \vert x_j )$ for all $i,k \in \mathcal{M}_j$ for all $j \in \mathcal{V}$. These estimates are erroneous in nature due to the non-zero false-alarm and misdetection probabilities in the system. As an example note that, to estimate $E[\gamma_i \vert x_j = 0]$ node $j$ needs to calculate an average over the samples of $\gamma_i$ received  when $x_j = 0$. However, node $j$ does not exactly know when $x_j = 0$ and it can only calculate $E[\gamma_i \vert \hat{x}_j = 0]$. When the sensing outcome indicates $\hat{x}_j = 0$, it may be, in fact, a misdetection and including the $\gamma_i$ sample in the averaging process may lead to a significant error in that case. Note that $\gamma_i$ samples can vary widely in size depending on the status of $x_j$. Similar errors can also occur due to the false-alarm incidents.  

Therefore, we need to enhance the quality of sensing in order to have high-quality estimates of the desired statistics which, in turn, are used to achieve a better sensing quality. Clearly, we are facing a loop here. Hence, we design an iterative learning-optimization loop to optimize the proposed message-passing algorithm. This loop works as an offline linear BP algorithm which operates on a window of $T$ samples of $\boldsymbol{\gamma}$. 

In the following, we use $\kappa$ (instead of $l$) to denote the iteration index of the offline linear BP algorithm in order to distinguish this iteration from the main linear BP algorithm. In addition, we use $\boldsymbol{\gamma}_{T}$ to denote a window of $T$ samples of $\boldsymbol{\gamma}$ stored for the proposed offline data processing and $\hat{\boldsymbol{x}}^{(\kappa)} \triangleq [\hat{x}_1^{(\kappa)}, ..., \hat{x}_N^{(\kappa)}]$ to denote the offline sensing outcomes, while we use $\boldsymbol{c}_j^{(\kappa)}$ and $\tau_j^{(\kappa)}$ to denote respectively the fusion coefficients and the detection threshold at node $j$ at (offline) iteration $\kappa$. For notational convenience, we do not explicitly show the timing of the samples here. However, it should be noted that by $\boldsymbol{\gamma}_{T}$ and $\hat{\boldsymbol{x}}^{(\kappa)}$ we are referring, respectively, to $T$ samples of $\boldsymbol{\gamma}(t)$ and $\hat{\boldsymbol{x}}^{(\kappa)}(t)$ corresponding to $t = 1,2, ..., T$. It is also worth noting the difference between $\hat{\boldsymbol{x}}^{(\kappa)}$ and $\hat{\boldsymbol{x}}$. Throughout the paper, we use $\hat{\boldsymbol{x}}$ to denote the sensing outcomes resulted from the main linear BP. However, $\hat{\boldsymbol{x}}^{(\kappa)}$ is used here to denote the offline sensing outcomes based on which the desired statistics are estimated. 


\captionsetup[table]{name= Algorithm}
\begin{table}[]
\centering
\caption{Blind Distributed Optimization of Linear BP by an Offline Adaptive Linear BP}
\label{tab:offlineBP}
\begin{tabular}{|l|l|l|l|l|}
\hline
\textbf{Adaptive Linear Belief Propagation  for Detector Optimization} \\
Input: $\boldsymbol{\gamma}_{T}$, $\boldsymbol{\tau}^{(0)}$, $\kappa_{\textup{max}}$, $\eta$ \\ 
Output: Near-optimal coefficients for linear BP
 \\ \thickhline 
 
1. \hspace{1mm} Let $\kappa \gets 0$ and initialize $\hat{\boldsymbol{x}}^{(0)}$ by comparing $\boldsymbol{\gamma}_{T}$ against $\boldsymbol{\tau}^{(0)}$ ; \\

2. \hspace{1mm} \textbf{while} $\kappa \le \kappa_{\textup{max}}$ \\

3. \hspace{3mm} \textbf{for} node $j \in \{1,2, ..., N\}$ \\

4. \hspace{5mm} Calculate $E[\gamma_i \vert \hat{x}_j^{(\kappa)}]$ and $\textup{cov}(\gamma_i, \gamma_k  \vert \hat{x}_j^{(\kappa)})$ for all $i,k \in \mathcal{M}_j$;  \\

5. \hspace{5mm} Solve \eqref{P4} to find coefficients $\boldsymbol{c}_j^{(\kappa)}$ and threshold $\tau_j^{(\kappa)}$;\\
 
 6. \hspace{3mm} \textbf{end}\\

7. \hspace{3mm} Run the resulting linear BP on $\boldsymbol{\gamma}_{T}$ to find  $\hat{\boldsymbol{x}}^{(\kappa+1)}$; \\

8. \hspace{3mm} $\kappa \gets \kappa + 1$; \\

9. \hspace{1mm} \textbf{end} \\

10.   \hspace{1mm} \textbf{for} node $j \in \{1,2, ..., N\}$ \\

11. \hspace{1mm} \hspace{2 mm} \textbf{for} $k \in \mathcal{M}_j$ \\

12. \hspace{1mm} \hspace{4 mm} \textbf{if} $c_j^{\textup{BP}}(k) / c_j^{(\kappa_{\textup{max}})}(k) > \eta$ \\

13. \hspace{1mm} \hspace{6mm} $c_j(k) \gets c_j^{\textup{BP}}(k)$; \\

14. \hspace{1mm} \hspace{4mm} \textbf{else}\\

 15. \hspace{1mm} \hspace{6mm} $c_j(k) \gets c_j^{(\kappa_{\textup{max}})}(k)$; \\
 
 16. \hspace{1mm} \hspace{4mm} \textbf{end}\\
 
 17. \hspace{1mm} \hspace{2mm} \textbf{end}\\
 
 18. \hspace{1mm} \textbf{end}\\

19. \hspace{1mm} Output $\boldsymbol{c}_j$ for $j \in 1,2,...N$;

\\\hline
\end{tabular}
\end{table}

 Algorithm \ref{tab:offlineBP} provides a pseudo-code of the proposed adaptation algorithm in which per-node computations are clarified. The adaptation is designed in a way that each node processes  samples of $\gamma_i$ only from its immediate neighbors. Hence, in the proposed distributed optimization no further information is used at each node other than those assumed available in \cite{Penna12}.

  The loop starts with initializing $\hat{\boldsymbol{x}}^{(0)}$ by the local sensing process on $\boldsymbol{\gamma}_{T}$. Specifically, for $t = 1, ..., T$,  initial offline sensing outcome $\hat{x}_j^{(0)}(t)$ is created at node $j$ by comparing $\gamma_j(t)$ against the corresponding local threshold $\tau_j^{(0)}$ calculated by \eqref{tau_0} which only requires the noise variance. The noise levels can be measured in practice \cite{Penna12}. At iteration $\kappa$, all $T$ samples of $\hat{\boldsymbol{x}}^{(\kappa)}$ are used to calculate $E[\gamma_i \vert \hat{x}_j^{(\kappa)}]$ and $\textup{cov}(\gamma_i, \gamma_k  \vert \hat{x}_j^{(\kappa)})$ for all $i,k \in \mathcal{M}_j$, $j = 1,2, ..., N$. These  estimates are then fed into \eqref{P4} to find the coefficients $\boldsymbol{c}_j^{(\kappa)}$ and thresholds $\tau_j^{(\kappa)}$. The resulting linear BP is then run over $\boldsymbol{\gamma}_{T}$ to find $\hat{\boldsymbol{x}}^{(\kappa+1)}$. 
The iteration stops after either a predetermined number of cycles or when there is no significant change in the updated parameters and statistics concerned. Lines 1 to 9 in Algorithm \ref{tab:offlineBP} describe this iteration.  
  
Due to the multipath fading or shadowing effects typically associated with the wireless environment, we may experience very low SNR levels at some sensing nodes. In such cases, the signals to be detected may be buried under a heavy noise and the channel estimation process may not be able to reliably capture the correlations between the desired random variables. In order to ensure that the system performance is not degraded in such conditions -- to below the performance level of the BP algorithm --  we compare the coefficients obtained by the adaptive linear BP against their counterparts in the linearized version of the main BP algorithm. If a coefficient obtained by the offline iteration is too small compared to its counterpart in the main BP, then that coefficient is not used in the adaptation and the one from the main BP is used instead. Lines 10 to 18 in Algorithm \ref{tab:offlineBP} illustrate this process. Specifically, if $c_j^{\textup{BP}}(k)/c_j^{(\kappa_{\textup{max}})}(k)$ is greater than a predefined value denoted by $\eta$, then $c_j^{\textup{BP}}(k)$ is used as the final fusion weight. Otherwise, $c_j^{(\kappa_{\textup{max}})}(k)$ is used. The final coefficients $\boldsymbol{c}_j$, $j=1,2,...N$, are then used in the main linear BP algorithm for spectrum sensing.

The proposed adaptation scheme enables us to rely on the effectiveness of the BP algorithm in capturing the interdependencies of the random variables in low SNR regimes while enhancing its performance by the linear fusion techniques whenever there is a room for such an enhancement. This is \emph{a hybrid design in which the two detection techniques work together.} The BP algorithm properly takes into account the outcome of the nodes in low SNR conditions while the linear fusion enhances the performance by injecting more information about the environment into the BP algorithm and this information is obtained from the nodes operating in better conditions. It is reasonable to assume that different nodes operate in different conditions due to the inherent spatial diversity in a wireless network. 
  
 It is worth noting that, the time is modeled as slotted in this paper and we refer to the proposed  adaptation process as offline to distinguish it from the main linear BP algorithm which is executed at the beginning of every time slot. Unlike the main linear BP iteration (or the legacy BP), Algorithm I is not run at every time slot. Instead, it is executed with a longer periodicity and in step with the changing characteristics of the radio environment. Since we assume that the statistical behavior of the environment changes slowly, the period over which Algorithm I is repeated is much longer than the time-slot duration. Specifically, if the spectrum sensing is performed at $t = nt_0$, $n=0,1,2,...$, with $t_0$ being the slot duration, then Algorithm I is executed at $t=kT_0$ for $k = 1,2, ...$, while $T_0/t_0 \gg 1$. Moreover, this adaptation is based on a window of data samples generated in the past compared to the data samples used for spectrum sensing. Hence, with respect to the spectrum sensing process, the fusion coefficients are adapted offline. 
  
\emph{Remark 6}: Since the statistics of the radio environment change slowly compared to the instantaneous status of $x_j$'s, the proposed adaptation is run far less frequently than is the main linear BP algorithm. Consequently, the proposed learning-optimization scheme does not lead to any significant increase in the computational complexity or communication overhead of the existing BP-based distributed detection methods. 

We do not use $\boldsymbol{\tau}^{(\kappa_{\textup{max}})}$  to set the detection thresholds in the main linear BP algorithm. The reason is that, to guarantee a certain level of the false-alarm or misdetection probability in the system we need a more accurate threshold adaptation mechanism. In the following, we propose a blind threshold adaptation scheme which guarantees the required performance.

\subsection{Detector Calibration}\label{subsec:Calibration}
As discussed in Section \ref{subsec:OfflineBP}, when we rely only on the detector outcome to derive the required conditional probabilities and statistics, we have to deal with inevitable errors. And, by evaluating \eqref{g} and \eqref{gApprox} we realize that the threshold values aimed at guaranteeing a certain false-alarm or detection probability appear to be too sensitive to these errors. Specifically, we have found through computer simulations that when $\eta_{j,v}(\boldsymbol{b})$, $\sigma_{j,v}^2$, and $\textup{Pr}\{\boldsymbol{x}_j = \boldsymbol{b} \vert x_j = v\}$ in \eqref{g} or $\hat{\eta}_{j,v}(\boldsymbol{b})$,  $\hat{\sigma}_{j,v}^2$, and $\textup{Pr}\{\boldsymbol{z}_j = \boldsymbol{b} \vert x_j = v\}$ in \eqref{gApprox} are estimated solely based on the detection outcomes, i.e., without  prior knowledge about the joint statistical behavior of $x_i$'s, the thresholds obtained by $g_j(\tau_j,0) \le \alpha$ or $g_j(\tau_j,1) \ge \beta$ cannot always guarantee the desired performance levels. However, they work well when the required statistics are available.  

Hence, we need a blind threshold adaptation method. We develop such an adaptation here by showing that the desired thresholds can be found through evaluating the outcomes of two sensing schemes which work in the same environment. It is clear that the sensing outcomes are always available and can be measured with no error. In this design, we use the fact that \emph{if the false-alarm (misdetection) probability of a sensing method can be set arbitrarily, that sensing method can be used to find a proper detection threshold for a better sensing scheme whose false-alarm (misdetection) probability needs to be below a certain level.} 
\begin{figure}[]
	\centering 
  \includegraphics[scale=0.45]{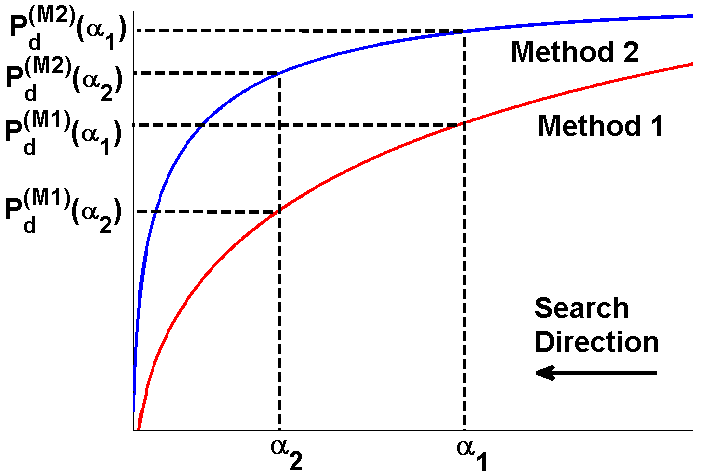} 
  \caption{ROC curves of two hypothetical sensing methods. Method 2 is a better detector but the false-alarm probability of Method 1 can be set arbitrarily.  }
  \label{fig2}
\end{figure}

 To clarify this observation, see Fig. \ref{fig2} in which the receiver operational characteristic (ROC) curves of two hypothetical sensing methods, Method 1 and Method 2, are depicted. Both methods are used to decide whether a binary-valued variable of interest, denoted here by $x$, equals 1. We assume that Method 2 exhibits a better  performance and that we can set the detection threshold in Method 1 such that its false-alarm probability is fixed at $\alpha_1$. Let $\tau_{\alpha_1}^{(\textup{M1})}$ denote that threshold. When $\tau_{\alpha_1}^{(\textup{M1})}$ is adopted for Method 1, the probability of observing 1 as the outcome of Method 1 can be stated as 
\begin{align}
\textup{Pr}&\{\hat{x}_{\textup{M1}}=1\} \nonumber \\
&= P_0\textup{Pr}\{\hat{x}_{\textup{M1}}=1 \vert x = 0\} + P_1 \textup{Pr}\{\hat{x}_{\textup{M1}}=1 \vert x = 1\} \nonumber \\
&= P_0 P_{\textup{f}}^{(\textup{M1})} + P_1P_{\textup{d}}^{(\textup{M1})} = P_0 \alpha_1 + P_1P_{\textup{d}}^{(\textup{M1})} \left(\alpha_1 \right)
\end{align}
where $\hat{x}_{\textup{M1}}$ denotes an estimation of $x$  given by Method 1 (i.e., the outcome of Method 1), $P_0 \triangleq \textup{Pr}\{x = 0\}$ and $P_1 \triangleq \textup{Pr}\{x = 1\}$, while $P_{\textup{f}}^{(\textup{M1})}$ and $P_{\textup{d}}^{(\textup{M1})}$ denote, respectively, the false-alarm and detection probabilities of Method 1. $P_{\textup{d}}^{(\textup{M1})} \left(\alpha_1 \right)$ denotes the detection probability of Method 1 when its false-alarm probability equals $\alpha_1$. In the following, we use similar notations regarding Method 2.

$\textup{Pr}\{\hat{x}_{\textup{M1}}=1\}$ can be measured in a simple fashion. We just need to count the number of times $\hat{x}_{\textup{M1}}$ equals 1 over a long enough period of time. 
Now, let $\tau^{(\textup{M2})}$ denote the detection threshold in Method 2. We are looking for $\tau_{\alpha_2}^{(\textup{M2})}$ at which $P_{\textup{f}}^{(\textup{M2})} = \alpha_2 \le \alpha_1$.  We start with $\tau^{(\textup{M2})} = 0$ and gradually increase the threshold level. At the starting point, Method 2 gives only 1 as its outcome. The reason is that when $\tau^{(\textup{M2})} = 0$, we have $P_{\textup{f}}^{(\textup{M2})} = P_{\textup{d}}^{(\textup{M2})} =1$. Hence, at the beginning $\textup{Pr}\{\hat{x}_{\textup{M2}}=1\} = P_0 + P_1 = 1$. It is also clear that, at the beginning we have $\textup{Pr}\{\hat{x}_{\textup{M2}}=1\} \ge \textup{Pr}\{\hat{x}_{\textup{M1}}=1\}$. 

By increasing $\tau^{(\textup{M2})}$ both values of $P_{\textup{f}}^{(\textup{M2})}$ and $P_{\textup{d}}^{(\textup{M2})}$ decrease reducing the gap between  $\textup{Pr}\{\hat{x}_{\textup{M2}}=1\}$ and $\textup{Pr}\{\hat{x}_{\textup{M1}}=1\}$. We stop the search when the two values coincide which happens when we have $P_{\textup{f}}^{(\textup{M2})} = \alpha_2$ and $\alpha_2 \le \alpha_1$. When the gap is zero we have $P_0 \alpha_2 + P_1P_{\textup{d}}^{(\textup{M2})} = P_0 \alpha_1 + P_1P_{\textup{d}}^{(\textup{M1})} \left(\alpha_1 \right)$. This leads to $P_{\textup{d}}^{(\textup{M2})} = \frac{P_0}{P_1} (\alpha_1 - \alpha_2) + P_{\textup{d}}^{(\textup{M1})} \left(\alpha_1\right)$. Therefore, now we have found a detection threshold which guarantees $P_{\textup{f}}^{(\textup{M2})} \le \alpha_1$ while $P_{\textup{d}}^{(\textup{M2})} \ge P_{\textup{d}}^{(\textup{M1})} \left(\alpha_1\right)$. 

The proposed threshold adaptation is based on measuring the difference in the behaviors of two sensing schemes operating in the same environment. Therefore, we refer to this process as the \emph{calibration of Method 2 by using Method 1}. 

\emph{Remark 7}: Assuming no prior knowledge about the environment means that none of the following probabilities are (perfectly) available: $P_0, P_1,  P_{\textup{d}}^{(\textup{M1})}, P_{\textup{f}}^{(\textup{M2})}$, and $P_{\textup{d}}^{(\textup{M2})}$. The proposed calibration finds the desired threshold level without the need for any of those parameters. 

In our spectrum sensing system, we have two options to use for calibrating the linear BP algorithm. The first option is the local sensing performed at each node. And the second one is the legacy BP-based distributed detection when the required false-alarm probability is small. In both of those detection schemes, we can guarantee a predefined level of false alarm only by knowing the noise level at each node. As mentioned before, the noise levels can be measured in practice. Since a linear BP with near-optimal coefficients exhibits a better performance level than the level shown by either of those methods, either of them can be used to calibrate the linear BP. We use the first option not to be limited to a small value of $\alpha$. 


\section{Convergence Conditions}\label{sec:Convergence}
An iterative message-passing algorithm which is run on a general factor graph does not necessarily converge to a fixed point. Even in case of the BP, when the graph contains loops, there is no guarantee, in general, that the beliefs will approximate the true marginals or even will converge to a fixed point at all. Furthermore, by using the proposed linear approximation in \eqref{m_kj_linear} we remove the nonlinear transfer functions and, consequently, increase the sensitivity of the nodes' outcomes to their input. Moreover, we increase the dynamic range of the values messages can attain. This change in the structure of the BP algorithm may deter its stability and convergence. However, by a careful design of the linear transformations, we can guarantee the convergence to a fixed point. In addition, by using the proposed optimizations we can enforce the algorithm to converge to a certain fixed point which leads to the desired performance. To realize this goal, we need to guarantee the so-called \emph{contraction mapping condition} \cite{Huang15, Mooij07} of the overall transformation obtained. 

 We collect all the messages $m_{i \rightarrow j}^{(l)}$, $(i,j) \in \mathcal{E}$ in the vector $\boldsymbol{m}^{(l)}$. Then, we can express the  message-update rule \eqref{m_kj_2} in matrix form as $\boldsymbol{m}^{(l)} = f(\boldsymbol{m}^{(l-1)})$ and the linear approximation as $\mathbf{m}^{(l)} \approx \boldsymbol{T} \boldsymbol{m}^{(l-1)} + \boldsymbol{\xi}$ where $\boldsymbol{T} = [T_{ij}]$ denotes an $|\mathcal{E}| \times |\mathcal{E}|$ matrix which contains $c_{ji}$'s while $\boldsymbol{\xi}$ denotes the offset in this transformation.  Assuming that the $i$th and $j$th elements of $\boldsymbol{m}^{(l)}$ are, respectively, $m_{k \rightarrow n}^{(l)}$ and $m_{p \rightarrow q}^{(l)}$, we have
 \begin{equation}\label{EQ36}
 T_{ij} = \frac{\partial m_{k \rightarrow n}^{(l)}}{\partial m_{p \rightarrow q}^{(l-1)}} = 
\left \{ \begin{matrix} c_{nk}, & p \in \mathcal{N}_k^n, q = k\\
 0, & \textup{otherwise}
 \end{matrix} \right.
 \end{equation}
 
Recall from Section \ref{sec:BP} that, we denote a general message-update rule as $\boldsymbol{m}^{(l)} = f^l(\boldsymbol{m}^{(0)})$ where $f : V \rightarrow V$. From \cite[Theorem 1]{Mooij07} we know that if there exist $0 \le K < 1$ such that $\|f(\boldsymbol{x})-f(\boldsymbol{y}) \| \le K \| \boldsymbol{x}-\boldsymbol{y}\|$ for any $\boldsymbol{x}, \boldsymbol{y} \in V$, then the sequence convergence is guaranteed. This is known as the contraction condition. Moreover, for a differentiable mapping $f$, for any point $\boldsymbol{z}$ on a segment between $\boldsymbol{x}$ and $\boldsymbol{y}$ denoted by $[\boldsymbol{x}, \boldsymbol{y}] \triangleq \{\theta \boldsymbol{x} + (1-\theta \boldsymbol{y}): \theta \in [0,1]\}$ we have $\|f(\boldsymbol{x})-f(\boldsymbol{y}) \| \le \| \boldsymbol{x}-\boldsymbol{y}\| \cdot \operatorname*{sup}_{\boldsymbol{z} \in [\boldsymbol{x}, \boldsymbol{y}]} \|f'(\boldsymbol{z})\|$. Therefore,  the following condition guarantees the convergence of a message passing algorithm defined by $f$
 \begin{equation}\label{EQ37}
 \operatorname*{sup}_{\boldsymbol{m} \in V} \|f'(\boldsymbol{m})\| <  1
 \end{equation}

  For simplicity, we work with the $\ell_{\infty}$ matrix norm $\| \cdot \|_{\infty}$. By using the the proposed linear approximation in \eqref{m_kj_linear}, we have $f(\boldsymbol{m}) = \boldsymbol{T} \boldsymbol{m} + \boldsymbol{\xi}$ which leads to $\|f'(\boldsymbol{m})\| = \|\boldsymbol{T}\|$.  Consequently, the contraction condition holds if  $\operatorname*{max}_{i} \sum_{j} |T_{ij}| < 1$
which, according to \eqref{EQ36}, is equivalent to $\operatorname*{max}_{k,n} \sum_{p} \sum_{q} \left |\frac{\partial m_{k \rightarrow n}}{\partial m_{p \rightarrow q}} \right | < 1$
which can be restated as $\operatorname*{max}_{k,n} \sum_{p \in \mathcal{N}_k^n } |c_{k,n}| < 1$
which is equivalent to $\operatorname*{max}_{k,n} (|\mathcal{N}_k|-1) |c_{k,n}| < 1$
where $|\mathcal{N}_k|$ indicates the number of neighbors of node $k$. 
Therefore, by imposing the following constraint on the proposed optimizations we guarantee the convergence of the proposed linear message-passing algorithm
\begin{eqnarray}\label{EQ42}
|c_{jk}| < \frac{1}{\operatorname*{max}_{n} |\mathcal{N}_n|-1},  \forall (j,k) \in \mathcal{E}
\end{eqnarray}
This is a convex constraint and does not add any significant challenge to the optimization problems at hand. 

It is worth noting that, \eqref{EQ42} is the last piece we need to develop an optimization framework for the BP-based distributed inference systems, which takes the advantages of both the BP and linear fusion. In particular, the BP algorithm is an effective approach to take into account the correlations between a group of spatially-distributed nodes while linear fusion is known to enable building low-complexity optimal decision variables based on the data exchanged between such nodes. By combining these capabilities, we develop a message-passing algorithm which takes into account the impact of pairwise correlations in accordance with the MRF model and is easy to optimize while its convergence is guaranteed.

\section{Impact of cycles} \label{sec:UTRW-BP}
Depending on the network topology, the factor graph may contain  cycles (loops) which might affect the quality of the marginals obtained by a message-passing algorithm. It is known that the presence of cycles in the graph can lead to the overconfidence in the beliefs which may degrade the system performance  \cite{Penna12}. The impact of cycles can be mitigated by using the so-called tree-reweigthed-BP (TRW-BP) algorithm \cite{Wymeersch12}. This method reduces the amount of information exchanged between the nodes to maintain an acceptable performance level when the factor graph contains cycles.  

In the TRW-BP the message from node $j$ to node $k$ is given by 
\begin{align}\label{EQ2}
\mu_{k \rightarrow j}^{(l)}(x_j)  \propto & \sum_{x_k}  \Bigg \{ \phi_k(x_k)  [\psi_{kj}(x_k,x_j)]^{\frac{1}{\rho_{kj}}}  \nonumber \\ 
&\left [\mu_{j \rightarrow k}^{(l-1)}(x_k)  \right ]^{\rho_{kj}-1}\prod_{n \in \mathcal{N}_k^j}\left [\mu_{n \rightarrow k}^{(l-1)}(x_k)  \right ]^{\rho_{nk}} \Bigg \}
\end{align}
while the beliefs are built as
\begin{equation}\label{EQ3} 
b_j^{(l)}(x_j) \propto \phi_{j}(x_j) \prod_{k \in \mathcal{N}_j}\left [\mu_{k \rightarrow j}^{(l)}(x_j) \right ]^{\rho_{kj}}
\end{equation}
where $\rho_{kj}$'s are referred to as the \emph{edge appearance probabilities} (EAP). In this method, we have $ 0<\rho_{kj} \le 1$ and the BP algorithm is a special case of the TRW-BP with $\rho_{kj} = 1$ for all $j,k$. Using smaller values for EAPs reduces the impact of cycles at the expense of cooperation between the sensing nodes. The performance of the TRW-BP can be optimized over the EAPs. However, this optimization grows rapidly in size with the number of nodes making it difficult to solve in a distributed manner. Assuming the same EAPs for all the links, i.e., $\rho_{j,k} = \rho$ for all $k,j$, shrinks the problem into a scalar optimization which can easily be solved in a distributed setting. The resulting method, which is referred to as uniform TRW-BP (UTRW-BP) algorithm, can achieve a good level of resistance against the impact of cycles in the graph. It is shown in \cite{Wymeersch12} that the optimal EAP is given by 
\begin{equation}
\rho = \min\{1,\frac{1}{2n_D}\}
\end{equation}
where $n_D$ is the average degree of the graph, i.e., the average number of connections of the nodes in the network. We know that, $n_D$ can be computed in a decentralized architecture by an
average consensus algorithm.   

Following the same approach as in Section \ref{sec:BP}, we express the messages and the decision variables achieved by the UTRW-BP as  
\begin{align} \label{m_kj_2_tr} 
m_{k\rightarrow j}^{(l)} &= \nonumber \\ 
&S\left(\frac{J_{kj}}{\rho}, ~\gamma_k + (\rho-1)m_{j\rightarrow k}^{(l-1)}+\rho \sum_{n\in {\cal N}_{k}^j} m_{n\rightarrow k}^{(l-1)}\right) \\
\lambda_j^{(l)} &= \gamma_j +\rho \sum_{k\in {\cal N}_{j}} m_{k\rightarrow j}^{(l)} 
\end{align}\\
which show that at every iteration the messages hit a damping factor before being combined together to from a new message or decision variable. And, the decision variable at node $j$ approaches approximately to 
\begin{align} \label{lambda_utrw}
\lambda_j  \approx \gamma_j &+ \rho\sum_{k \in \mathcal{N}_j}c_{jk}\gamma_k + \rho^2\sum_{k \in \mathcal{N}_j}\sum_{n \in \mathcal{N}_k^j}c_{jk}c_{kn} \gamma_n  \nonumber\\
  &+ \rho^3\sum_{k \in \mathcal{N}_j}\sum_{n \in \mathcal{N}_k^j}\sum_{m \in \mathcal{N}_n^k}c_{jk}c_{kn}c_{nm} \gamma_m + ... 
\end{align}
where $c_{jk} = \frac{(e^{2J_{kj}/\rho}-1)}{(1+e^{J_{kj}/\rho})^2}$. 

From \eqref{lambda_utrw} we obtain some interesting observations. In particular, we observe a stronger decaying effect by distance from node $j$ than we had before. As we saw in \eqref{lambda_j_linear}, since $|c_{jk}| < 1$, the impact of other nodes on $\lambda_j$ decreases by the distance of those nodes from node $j$. Now, this decaying is stronger and in favor of the one-hop neighbors even more than before. Specifically, if $\gamma_i$ arrives at node $j$ through a path of length $L$, then it is multiplied by $\rho^L$ in the data-fusion process. Consequently, in the presence of loops, i.e., when there is more than one path from node $i$ to node $j$, the impact of $\gamma_i$ on $\lambda_j$ received through the shortest path is amplified compared to the ones received through the other paths. In other words, in the UTRW-BP the decision variable at a node is built mainly by the data received through the tree sub-graph(s) connecting that node to all the other nodes via the shortest paths.  Hence, $\rho$ reduces the overconfidence in nodes by alleviating the impact of cycles. 

We can accommodate this decaying effect in the proposed linear BP scheme. Note that, for a fixed false-alarm probability $P_{\textup{f}}^{(j)}$ the detection probability $P_{\textup{d}}^{(j)}$ is positively homogeneous of degree zero with respect to $\boldsymbol{c}_j$. To show this we set the detection threshold in \eqref{gApprox} as 
 \begin{equation}
 \tau_j = \mathfrak{q}_\alpha\tilde{\sigma}_{j,0}(\boldsymbol{b}) + \tilde{\eta}_{j,0}(\boldsymbol{b}) 
 \end{equation}
where $\mathfrak{q}_\alpha \triangleq Q^{-1}(\alpha)$. Consequently, $P_{\textup{f}}^{(j)} \approx \alpha$ and 
 \begin{equation}
P_{\textup{d}}^{(j)} \approx \sum_{\boldsymbol{b} \in \{0,1\}^{\left|\mathcal{N}_j\right|}} Q \left(\frac{\mathfrak{q}_\alpha\tilde{\sigma}_{j,0}(\boldsymbol{b}) - \Delta \tilde{\eta}_j(\boldsymbol{b})}{\tilde{\sigma}_{j,1}(\boldsymbol{b})} \right )p_{\tilde{\boldsymbol{x}}_{(j)} |x_j}(\boldsymbol{b}|1)
 \end{equation}
 where $\Delta \tilde{\eta}_j \triangleq \tilde{\eta}_{j,1}(\boldsymbol{b}) - \tilde{\eta}_{j,0}(\boldsymbol{b})$. 
 
Now we see that by using $\rho\boldsymbol{c}_j$ instead of $\boldsymbol{c}_j$ as the fusion vector at node $j$, the detection probability $P_{\textup{d}}^{(j)}$ remains unchanged. The reason is that, the variables $\tilde{\sigma}_{j,0}, \tilde{\sigma}_{j,1}$ and $\Delta \tilde{\eta}_j$ are all multiplied by $\rho$, thus, canceling the effect of this scaling on $P_{\textup{d}}^{(j)}$. Since the detection probability is a monotonically increasing function of the false-alarm probability, the global optimum in \eqref{P4} occurs when $P_{\textup{f}}^{(j)} = \alpha$. Therefore, given $\boldsymbol{c}_j$ as a solution for \eqref{P4}, its scaled version $\rho\boldsymbol{c}_j$ is also a global optimum. Hence, by using $\rho c_{jk}$ in \eqref{m_kj_linear} we obtain a similar decaying effect to the one observed in UTRW-BP without altering the proposed optimization framework. Note also that, using the deflection coefficient as the objective function in the proposed distributed optimization does not affect this argument since the deflection coefficient is  a homogeneous function of $\boldsymbol{c}_j$ as well.
 
As the final point, note that when the number of iterations is small there is no need to be concerned of the impact of cycles \cite{Penna12}. 

\section{Numerical Results}\label{sec:Simulations}
\begin{figure}[]
	\centering 
  \includegraphics[scale=0.40]{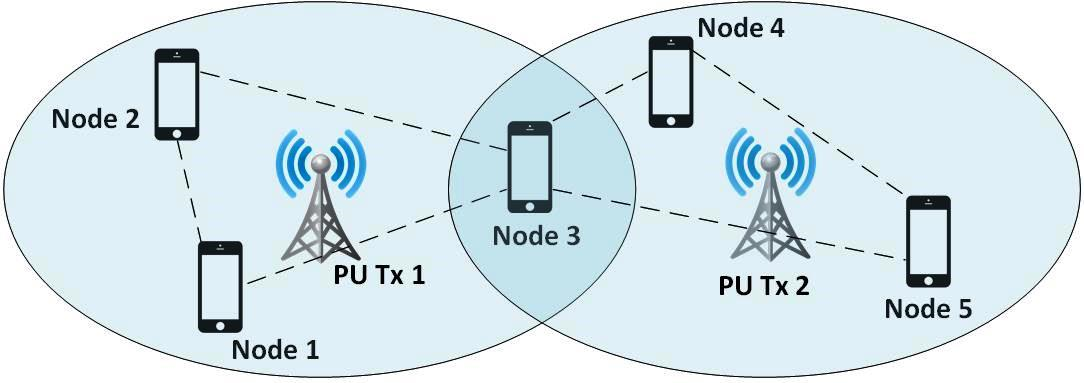} 
  \caption{The network configuration considered in the simulations. Five sensing nodes cooperate to find transmission opportunities within the spectrum bands allocated to two primary transmitters. The dashed lines depict the links between the sensing nodes through which the distributed detection is conducted.    }
  \label{fig:network}
\end{figure}

Fig. \ref{fig:network} depicts the network configuration used in our simulations. We consider two PU transmitters and five sensing nodes as SUs.  Nodes 1 and 2 are located within the range of PU transmitter 1, nodes 4 and 5 are located within the range of PU transmitter 2, while node 3 is located within the range of the both PU transmitters. These nodes exchange messages with each other over a set of wireless links, depicted as the dashed lines in the figure, to conduct a distributed detection scheme through which the transmission opportunities are discovered within the spectrum bands allocated to the PUs.  

Energy detection is used as the local detection method at each node with $K = 100$ samples to estimate the received signal energy. The SNR levels of the signals received at nodes 1 and 5 are -5 dB, at nodes 2 and 4 are -8 dB and at node 3 are -10 and -10 dB. We realize a spatially-correlated occupancy pattern by making the primary transmitters exhibit correlated random on and off periods. This is an extension of the simulation scenario in \cite{Penna12} where one of the PU transmitters is constantly on while the other one is off all the time. The channel coefficients are  constant during the sensing period as in \cite{Penna12}. The number of iterations is 3 for all the message-passing algorithms considered in this section. In addition, the offline BP algorithm is repeated only 4 times, i.e., $\kappa_{max} = 4$. Each data point in the simulation results is obtained by averaging the detection outcomes over 20,000 time slots. A window of $T = 2000$ time slots is used to train all the message-passing algorithms.

\begin{figure}[]
	\centering 
  \includegraphics[scale=0.35]{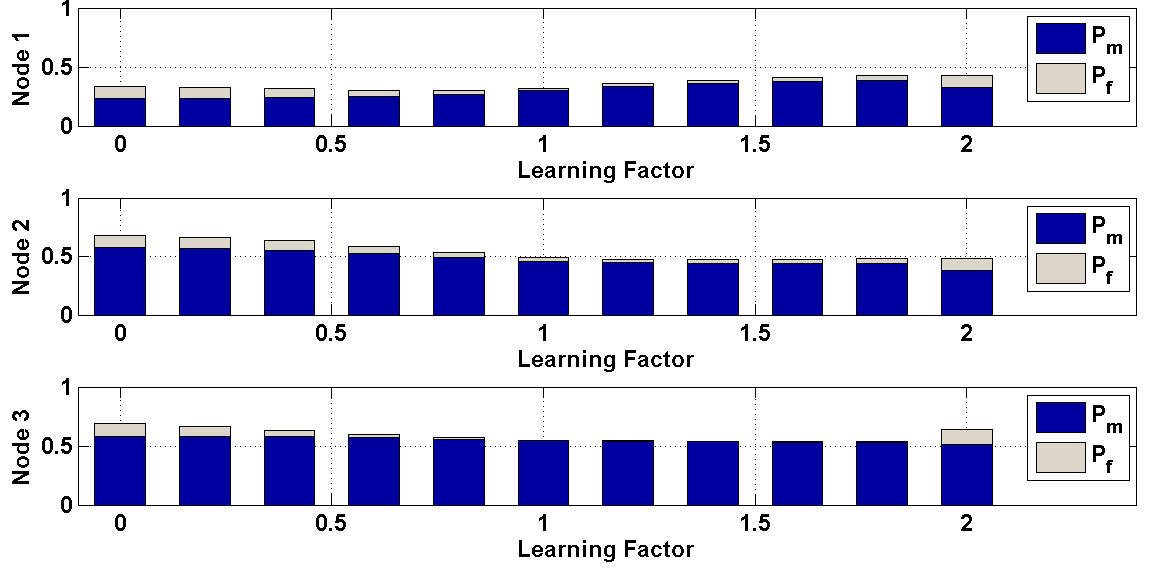} 
  \caption{Different levels of system performance, in terms of the false-alarm  and misdetection probabilities, denoted $P_{\textup{f}}$ and $P_{\textup{m}}$ respectively, are observed for different values of the learning factor $\zeta$.  }
  \label{fig:learnFactor}
\end{figure}

Fig. \ref{fig:learnFactor} depicts the system performance in terms of the false-alarm and misdetection probabilities vs. the learning factor in \eqref{lrnFact}. We observe different performance levels for different values of the learning factor, which indicates the need for optimization. To cover a wide range of possible performance levels of the BP algorithm, we adopt three values as $\zeta = 0.2, 0.4, 1$. As reported in \cite{Penna12}, the BP algorithm works better when the learning factor is small. However, note that how to determine the best value for $\zeta$ is unknown. For the sake of brevity, we only focus on nodes 1, 2, and 3.  Nodes 4 and 5 experience similar conditions as those in which nodes 1 and 2 operate. The optimal linear fusion scheme is realized in this section by solving \eqref{P2} via the sequential optimization method described in \cite{Quan09}.

\begin{figure}[]
	\centering 	
	\begin{subfigure}{.45\textwidth}
	  \centering 
  \includegraphics[scale=0.53]{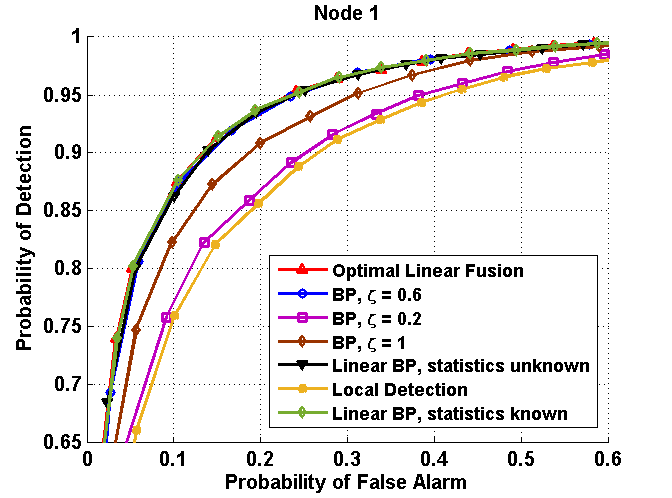} 
  \caption{Detection performance of node 1.}
  \label{fig:rocNode1} 
  \end{subfigure}  
  \begin{subfigure}{.45\textwidth}
    \centering 
  \includegraphics[scale=0.53]{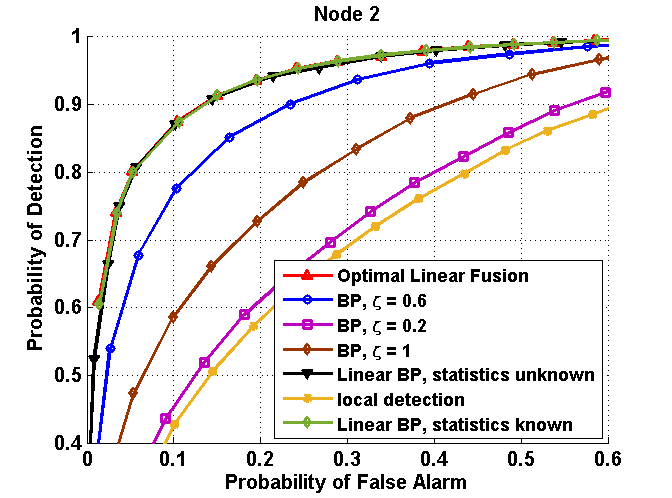} 
  \caption{Detection performance of node 2.}
  \label{fig:rocNode2} 
  \end{subfigure}  
  \caption{Comparison of the different distributed detection methods discussed in this paper. The proposed linear BP obtains the performance of optimal linear fusion without having the statistics which characterize the radio environment.}
\end{figure}

We use ROC curves to demonstrate  and compare the detection performance achieved by the different methods. In Fig. \ref{fig:rocNode1} we see the performance of node 1 which operates under a relatively better SNR regime than those experienced by nodes 2 and 3. The performance of the proposed linear BP along with those of the optimal linear fusion and the local detection are shown for comparison. We first observe that the performance of the BP algorithm improves by increasing the value of $\zeta$ from 0.2 to 0.6 and then degrades by a further increase in $\zeta$. Moreover, we observe that both the BP with $\zeta = 0.6$ and the proposed linear BP achieve the performance of the optimal linear fusion scheme without any prior knowledge available concerning the radio environment. In comparing these performance results it should be noted that, firstly, $\zeta = 0.6$ is found by a numerical search and secondly, as it is shown later in this section, it fails to always guarantee the performance constraint on the system false-alarm probability. The proposed method achieves the optimal performance blindly while guaranteeing that the required constraint on the false-alarm probability is always met. 

Fig. \ref{fig:rocNode2} depicts the detection performance obtained at node 2. Here we have a clear performance gap which illustrates that the proposed method outperforms the conventional BP algorithm. Therefore, we can see that when operating under a lower SNR level (than the one observed in node 1), the proposed linear BP achieves the detection rate of the optimal linear fusion while the BP algorithm falls behind. This can be attributed to the better learning capability and also to the better combination of the local test summaries in the proposed linear BP algorithm. Specifically, we know that a properly designed linear fusion scheme provides a constructive discrimination \cite{Abdi17} on the local sensing outcomes when they are combined together to build a decision variable. This discrimination alleviates the degrading effect of the local sensing results obtained in low SNR regimes by emphasizing the effect of the ones created by the more reliable nodes. In Fig. \ref{fig:rocNode2}, it appears that such discrimination is better realized in the proposed linear BP algorithm compared to the conventional BP. Note that, both methods use the same data which is obtained through the exchange of information between the one-hop neighbors in the network. 

\begin{figure}[]
	\centering 
	\begin{subfigure}{.45\textwidth}
	\centering
  \includegraphics[scale=0.53]{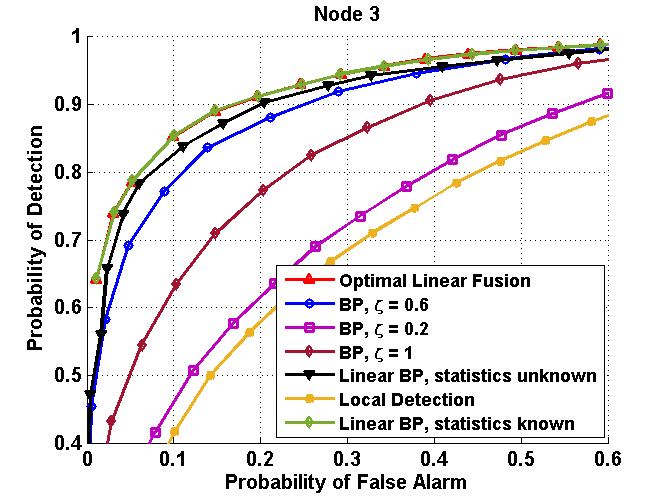} 
  \caption{Detection performance of node 3.}
  \label{fig:rocNode3}
  \end{subfigure}
  \begin{subfigure}{.45\textwidth}
	\centering
  \includegraphics[scale=0.53]{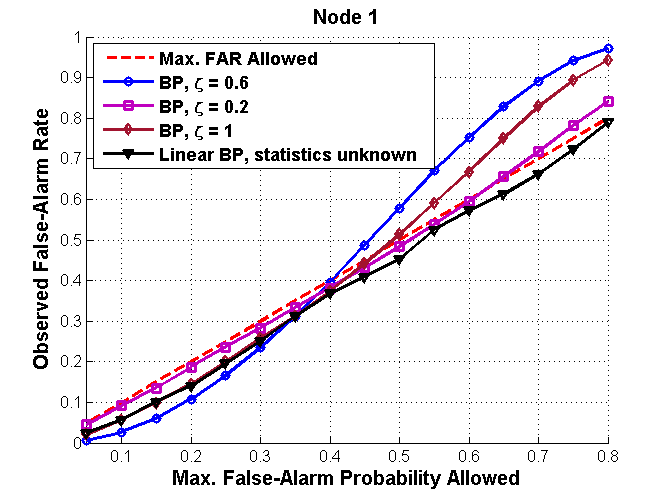} 
  \caption{FAR observed at node 1.}
  \label{fig:FARs}
  \end{subfigure}
    \caption{Detection performance at node 3 and the result of different threshold adaptation schemes at node 1. The linear BP algorithm outperforms the BP algorithm and exhibits near-optimal performance and the proposed threshold adaptation scheme guarantees the FAR fall below the predefined threshold. }
\end{figure}

Fig. \ref{fig:rocNode3} depicts the detection performance of node 3. Again, we  see that the optimal linear BP outperforms the conventional BP algorithm. And, the linear BP nearly attains the detection rate of the optimal linear fusion scheme without having any prior knowledge about the radio environment. 

By comparing the performance gap between the local detection scheme and all other distributed detection methods considered in this section we see that the impact of a distributed detection scheme is more significant when the nodes operate in low SNR regimes and that is the situation in which the proposed method offers a higher performance gain than what is achieved by the conventional BP algorithm. In other words, in a network of distributed sensors, when the SNR levels are low we have a stronger need to establish some sort of cooperation between the sensing nodes to achieve more-reliable sensing results and the linear BP algorithm offers a more significant gain, over the other methods, in such a condition. 

The performance of the proposed detector calibration method is depicted in Fig. \ref{fig:FARs} where the false-alarm rate (FAR) observed is depicted vs. the constraint imposed on the system false-alarm probability. We see that when the detection threshold is determined by the proposed calibration, the FAR is always guaranteed to fall below the predefined upper bound. However, when the conventional BP is used with the detection threshold calculated by \eqref{tau_0}, it fails to achieve the required performance guarantees when the constraint is not small enough. Equivalently, Fig. \ref{fig:FARs} shows that, when \eqref{tau_0} is used, to achieve the required performance guarantee for an arbitrary $\alpha$, we have to keep the learning factor $\zeta$ relatively small. However, as we saw previously by inspecting the ROC curves, when $\zeta$ is small the detection performance is far from optimal. Hence, the proposed linear BP algorithm outperforms the existing BP algorithm when the FAR of the system is required to be controlled. 

It is worth noting that, the use of \eqref{Penna'sED} is inspired by \eqref{gamma_j} which is derived assuming that if $x_j = 1$, then $\boldsymbol{y}_j$ is a vector of Gaussian random variables. However, this is not always the case. From \eqref{y_j} we see that, when there exist more than one PU transmitter, as in our network configuration, $\boldsymbol{y}_j$ cannot be assumed necessarily to follow a Gaussian distribution. Consequently, the violation of the FAR constraint when $\tau_0$ is used appears to be due to the error in modeling the statistical behavior of $\gamma_j$. The proposed calibration method works based on modeling the behavior of $\gamma_j$, given $x_j = 0$, by a Gaussian distribution which is, in fact, the case. Therefore, the proposed system is capable of meeting the required FAR constraint. 
   
As the final point, the better performance of the proposed threshold adaptation scheme does not stem from the linearity of the proposed message-passing algorithm since the calibration method proposed does not depend on the behavior of the transfer functions affecting the messages. For the sake of brevity, Fig. \ref{fig:FARs} depicts only the FAR observed at node 1. All other nodes show a similar performance.  


\section{Conclusions} \label{sec:conclusions}
We demonstrated that a BP algorithm in a network of distributed agents can be viewed or approximated as a distributed linear data-fusion scheme. Accordingly, we addressed two particular issues in using the BP algorithm for distributed detection: First, the lack of knowledge about the statistical behavior of the BP messages exchanged between the nodes and second, the lack of a proper threshold adaptation mechanism in a BP-based detection scheme. 

Regarding the first issue, our analysis and simulation results demonstrate that, when the local sensing results follow the Gaussian distribution, 
\begin{itemize}
\item In a centralized configuration and when the first- and second-order statistics of the local detection results of all nodes are available at the FC, a linear fusion of all the local detection results, realized by \eqref{P1}, can obtain the desired performance and there is no need for an iterative message-passing algorithm. 

\item In a decentralized configuration and when at each node the first- and second-order statistics of the local detection results of its one-hop neighbors are available, the desired performance can nearly be obtained via a linear message-passing algorithm realized by \eqref{P3}. 

\item In a decentralized configuration and when there is no statistics available a priori, the desired performance can be obtained via a linear message-passing algorithm in which the weighting coefficients are obtained blindly via the offline adaptive linear BP algorithm described in Section \ref{subsec:OfflineBP}. 
\end{itemize}

And, regarding the second issue, we have shown that in the proposed linear BP algorithm, 
\begin{itemize}
\item The detection threshold to guarantee a certain false-alarm (or misdetection) probability can be found analytically at each node when the first- and second-order statistics of the one-hop neighbors of that node are available.
 
\item The system false-alarm (or misdetection) probability can be guaranteed to fall below a certain level by the calibration method proposed in Section \ref{subsec:Calibration} without the need for any statistics describing the behavior of the local test summaries or the wireless environment. 
\end{itemize}

\bibliographystyle{IEEEtran}
\bibliography{IEEEabrv,Bibliogeraphy_ICLAS}

\end{document}